\documentclass[12pt]{article}
\usepackage{amsmath,amssymb,amsthm,amscd} 
\usepackage{hyperref} 
\usepackage[all]{xy}
\usepackage{graphicx}
\usepackage[parfill]{parskip}
\usepackage[usenames]{color}
\usepackage{tikz}
\usepackage[nottoc]{tocbibind}
\usetikzlibrary{shapes.geometric}
\usetikzlibrary{snakes}
\usepgflibrary{arrows}
\numberwithin{equation}{section}
\numberwithin{table}{section}

\newdimen\tableauside\tableauside=1.0ex
\newdimen\tableaurule\tableaurule=0.4pt
\newdimen\tableaustep
\def\phantomhrule#1{\hbox{\vbox to0pt{\hrule height\tableaurule width#1\vss}}}
\def\phantomvrule#1{\vbox{\hbox to0pt{\vrule width\tableaurule height#1\hss}}}
\def\sqr{\vbox{%
  \phantomhrule\tableaustep
  \hbox{\phantomvrule\tableaustep\kern\tableaustep\phantomvrule\tableaustep}%
  \hbox{\vbox{\phantomhrule\tableauside}\kern-\tableaurule}}}
\def\squares#1{\hbox{\count0=#1\noindent\loop\sqr
  \advance\count0 by-1 \ifnum\count0>0\repeat}}
\def\tableau#1{\vcenter{\offinterlineskip
  \tableaustep=\tableauside\advance\tableaustep by-\tableaurule
  \kern\normallineskip\hbox
    {\kern\normallineskip\vbox
      {\gettableau#1 0 }%
     \kern\normallineskip\kern\tableaurule}%
  \kern\normallineskip\kern\tableaurule}}
\def\gettableau#1 {\ifnum#1=0\let\next=\null\else
  \squares{#1}\let\next=\gettableau\fi\next}

\newcommand{\be}{\begin{equation}}
\newcommand{\ee}{\end{equation}}
\newcommand{\ben}{\begin{equation*}}
\newcommand{\een}{\end{equation*}}
\newcommand{\diag}{\mathrm{diag}}

\newcommand{\re}{\mathrm{Re}}
\newcommand{\im}{\mathrm{Im}}

\newcommand{\ttd}{(t,t_D)}

\def\beq{\begin{eqnarray}}
\def\eeq{\end{eqnarray}}
\def\ba{\begin{eqnarray}}
\def\ea{\end{eqnarray}}
\def\ban{\begin{eqnarray*}}
\def\ean{\end{eqnarray*}}

\def\ep1{\epsilon_1}
\def\eps2{\epsilon_2}

\newcommand{\Ztop}{\mathrm{Z_{top}}}
\newcommand{\Znek}{\mathrm{Z_{nek}}}
\newcommand{\Zop}{\mathrm{Z_{open}}}

\newcommand{\IZ}{\mathbb{Z}}
\newcommand{\IC}{\mathbb{C}}
\newcommand{\IP}{\mathbb{P}}

\newcommand{\IR}{\mathbb{R}}

\newcommand{\IH}{\mathbb{H}}

\newcommand{\xf}{X_{\rm{fid}}}
\newcommand{\zf}{\mathrm{Z_{fid}}}
\newcommand{\zs}{\mathrm{Z_{strip}}}

\newcommand{\tr}{\mathrm{Tr }}
\newcommand{\Tr}{\mathrm{Tr }}

\newcommand{\nn}{\nonumber}

\newcommand{\cW}{{\cal W}}
\newcommand{\cN}{{\cal N}}
\newcommand{\cM}{{\cal M}}
\newcommand{\cS}{{\cal S}}

\newcommand{\cO}{{\cal O}}
\newcommand{\cC}{{\cal C}}

\newcommand{\cF}{{\cal F}}

\newcommand{\cK }{{\cal K}}

\newcommand{\Vc}{V}
\newcommand{\Vt}{V}

\newcommand{\del}{\partial}

\newcommand{\aaa}{e_1(\tau_{uv})} 
\newcommand{\bbb}{e_2(\tau_{uv})} 
\newcommand{\ccc}{e_3(\tau_{uv})} 
\newcommand{\AAA}{e_1(\tau)} 
\newcommand{\BBB}{e_2(\tau)} 
\newcommand{\CCC}{e_3(\tau)} 

\def\ep1{\epsilon_1}
\def\eps2{\epsilon_2}
\def\ha{h_0}

\def\hd{h_3}

\begin{document}

\begin{titlepage}

\newcommand{\HRule}{\rule{\linewidth}{0.2mm}} % Defines a new command for the horizontal lines, change thickness here

\center % Center everything on the page
 
%----------------------------------------------------------------------------------------
%	HEADING SECTIONS
%----------------------------------------------------------------------------------------

%\large{Laboratoire de Physique Th\'eorique}\\[0.5cm] % Name of your university/college
\large{LABORATOIRE DE PHYSIQUE TH\`EORIQUE} \\
\large{DE L'\'ECOLE NORMALE SUP\'ERIEURE}

\begin{figure*}[h]
\centering
\includegraphics[width=0.2\textwidth]{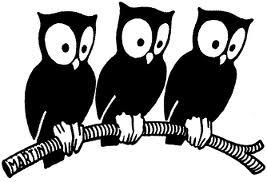}
\end{figure*}
%\include
%{\Large Th\`ese d'habilitation \`a diriger des recherches}\\[0.5cm] % Major heading such as course name
{\Large TH\`ESE D'HABILITATION \`A DIRIGER DES RECHERCHES} \\[0.5cm] 
{\large pr\'esent\'ee par}\\[0.5cm] % Major heading such as course name
{\Large \bf Amir-Kian Kashani-Poor}\\[0.5cm] % Minor heading such as course title

%----------------------------------------------------------------------------------------
%	TITLE SECTION
%----------------------------------------------------------------------------------------

\HRule \\[0.7cm]
{ \huge \bfseries Computing $\mathbf{Z_{top}}$}\\[0.4cm] % Title of your document
\HRule \\[1.5cm]
 
%----------------------------------------------------------------------------------------
%	AUTHOR SECTION
%----------------------------------------------------------------------------------------

{\large Soutenue le 28 mai 2014}\\[1.4cm]

{\large Composition du jury :}\\[0.6cm]

\begin{minipage}{0.9\textwidth}
\begin{tabbing}
M. Ignatios Antoniadis \quad \=(Rapporteur)\\
M. Costas Bachas\\
M. Atish Dabholkar  \\
M. Albrecht Klemm  \\
M. Boris Pioline  \>(Rapporteur) \\
M. Alessandro Tanzini \>  (Rapporteur)\\
M. Jean-Bernard Zuber \\
\end{tabbing}
\end{minipage}

\vfill % Fill the rest of the page with whitespace

\end{titlepage}

\mbox{}
\thispagestyle{empty}
\newpage
\pagenumbering{Roman}
\hspace{4cm}
\section*{Abstract}
The topological string presents an arena in which many features of string theory proper, such as the interplay between worldsheet and
target space descriptions or open-closed duality, can be distilled into computational techniques which yield results beyond perturbation
theory. In this thesis, I will summarize my research activity in this area. The presentation is organized around computations of the topological string partition function $\Ztop$ based on various perspectives on the topological string.

\hspace{2cm}

\section*{R\'esum\'e}
La corde topologique \'etablit un contexte dans lequel beaucoup de caract\'eristiques de la th\'eorie des cordes compl\`ete, comme les aspects compl\'ementaires de la description par la surface d'univers et l'espace cible, ou la dualit\'e entre corde ouverte et corde ferm\'ee, m\`enent \`a des techniques de calcul qui vont au-del\`a de la th\'eorie perturbative. Ce m\'emoire r\'esume mes activit\'es de recherche dans ce domaine. La pr\'esentation est organis\'ee autour des calculs de la fonction de partition de la corde topologique $\Ztop$ bas\'es sur des perspectives diverses sur la corde topologique.

\newpage
\null 
\newpage

\tableofcontents
\clearpage\null   
%\thispagestyle{empty} 
%\newpage
\pagenumbering{arabic}
\setcounter{page}{1}

\section{Introduction: What is $\Ztop$?} \label{s:intro}
In the course of the last century, all fundamental forces underlying the dynamics of nature were quantized, and the resulting theories convincingly tested experimentally, save one: the force of gravity. That quantizing gravity requires new ideas comes as no surprise, given the necessity, as demonstrated by Einstein's theory of general relativity, of elevating space-time, in the process, from the spectator role it plays for the other fundamental forces to a dynamical participant. String theory is work in progress towards such a quantum theory of gravity. It indeed introduces a plethora of new ideas to the task. Unifying these into a concise theoretical framework is an important goal of research in the field. 

The original formulation of string theory, perturbative string theory, nudges space-time out of its spectator role by elevating space-time coordinates to dynamical fields. In contradistinction to conventional field theories, in which the dynamical quantities are functions of space-time coordinates, perturbative string theory takes its fundamental dynamical object, the eponymous string, as a starting point. Its worldsheet, the two dimensional surface the one dimensional string traces out while propagating in time, is the domain of the field theory underlying perturbative string theory. The fields encode how this worldsheet is mapped into space-time. Quantization of these maps gives rise to the excitations which we observe as matter particles and force carriers, including the graviton, the carrier of the gravitational force. Space-time is the target of these maps, and the conventional field theory approach is therefore referred to as a target space description from this perspective.

A fundamental challenge of the perturbative string is indicated by the qualifier {\it perturbative}. Scattering amplitudes of the excitations of the theory are obtained in a perturbation series in the gravitational coupling, which is encoded in the string coupling constant $g_s$. The power of $g_s$ is a measure of how many times a string splits and joins as it propagates through space-time while realizing a certain scattering process. As the worldsheet theory is in fact conformal, the worldsheet can be mapped to Riemann surfaces with punctures, such that the splitting and joining frequency is measured by the genus of the Riemann surface. Successive terms in the perturbation series in $g_s$ are thus obtained by repeating the scattering calculation for worldsheets of higher and higher genera. The scattering amplitude thus obtained should be the perturbative, probably only asymptotic, approximation to an underlying exact result. Finding a formulation of string theory in which this exact result can be formulated as an observable, and which reduces to the perturbative string in a perturbative approximation, is a fundamental problem of the field, dubbed {\it the search for a non-perturbative completion of string theory}.

Topological string theory describes a sub-sector of the full physical string theory, in essence by only retaining the zero-modes in the physical spectrum. The resulting simplification enhances our mathematical and computational control of the theory. The interest in the topological string is fed from various sources. Most immediately, the topological string is directly related to the physical string on space-times of the form $\IR^{3,1} \times X$, with the first factor the four dimensional space-time of our quotidian experience, and $X$ a six dimensional Calabi-Yau manifold. It computes contributions to the effective space-time action of this theory and is used as a tool to explore the low energy manifestations of string theory. Mathematicians are interested in the topological string as a technique for studying families of Calabi-Yau manifolds, and associated structures, extending to integrability, number theory, and beyond. Finally, the study of topological string theory is fuelled by the ambition to extract structural lessons applicable to the full physical string, in particular concerning the question of its non-perturbative completion. 

Just as perturbative string theory, the topological string in its original formulation is defined genus by genus, and gives rise to functions $F_g$ of geometric parameters $t$ characterizing its target space $X$, a Calabi-Yau manifold. From the A-model point of view, these take the schematic form
\be
F_g = \sum_{\beta \in H_2(X)} N_{g,\beta} \, e^{-\beta \cdot t} \,.
\ee
The $N_{g, \beta}$ are rational numbers called Gromov-Witten invariants. Integers can be extracted from these that count, in an appropriate sense, the number of holomorphic maps from a genus $g$ Riemann surface to the target space $X$, such that the image has homology class $\beta$. These maps are the zero modes of the worldsheet fields alluded to above that embed the physical string into space-time. Assembling the topological string amplitudes $F_g$ into a formal generating function gives rise to the topological string partition function
\be  \label{ztop_basic}
\Ztop = \exp \sum_{g=0}^\infty F_g\,  g_s^{2g-2}  \,.
\ee
The name {\it partition function} indicates that, just as in the physical string, we expect an underlying theory within which $\Ztop$ is elevated beyond its perturbative definition. Unlike the situation in the physical string, the quantity $\Ztop$ is understood sufficiently well in a variety of circumstances to permit its study beyond perturbation theory. This is the topic of this thesis.

We will be focussing throughout on non-compact manifolds $X$ (though the methods of section \ref{section:hae} apply equally well to compact Calabi-Yau manifolds). In this decompactification limit, gravity becomes arbitrarily weak, which is why gauge theories will feature prominently in our discussion. It is perhaps disappointing to begin the journey at such a distance from the proclaimed goal, a quantum theory of gravity. As re-assurance, we can offer that the fundamental problem of finding a non-perturbative completion to a perturbative worldsheet theory persists to this level of simplification -- as consolation, that we will encounter much beautiful and intricate structure along the way.

\newpage

\section{$\Ztop$ via geometric transitions} \label{s:geom_tran}
In this section, we will discuss how a target space description of the open topological string gives rise to an algorithm for computing $\Ztop$ non-perturbatively in $g_s$ on any toric Calabi-Yau manifold $X$. The expressions obtained in this formalism are closely related to the partition function in Gopakumar-Vafa form,
\be  \label{ztop_gv}
\Ztop = \exp \left[ \sum_{g=0}^\infty \sum_{\beta \in H_2(X)} \sum_{n=1}^\infty \frac{n^g_\beta}{n} \left( 2 \sin \frac{n g_s}{2} \right)^{2g -2} e^{- n\beta \cdot t}  \right]\,,
\ee
which is derived in \cite{Gopakumar:1998ii,Gopakumar:1998jq} by identifying $\Ztop$ with a space-time index on the spectrum of the physical theory. At a given $\beta$, only finitely many of the Gopakumar-Vafa invariants $n^g_\beta$ are non-vanishing. Hence, presented in this form, the coefficients of $\Ztop$ in an expansion in K\"ahler parameters $n\beta \cdot t$ are analytic functions in $g_s$. In contrast, much of the structure of $\Ztop$ as a function of the K\"ahler parameters is hidden in the series expansion over $\beta$. A central focus in this section will be the partial resummation of this series. In particular, this will allow the proof of the equality of $\Ztop$ on certain geometries with the Nekrasov partition function of $\cN=2$ gauge theories, which will play a pivotal role in furthering our understanding of $\Ztop$ in section~\ref{s:agt}.

\subsection{Chern-Simons theory as the target space description of the open topological string}
In \cite{Witten:1992fb}, Witten identified $U(N)$ Chern-Simons theory on a compact 3-manifold $M$ as the target space description of the open topological string 
A-model on $T^* M$, the cotangent space of $M$, with $N$ A-branes wrapping the Lagrangian submanifold $M$ of $T^* M$. A simple argument \cite{Witten:1992fb} proves that no holomorphic maps $\phi: \Sigma \rightarrow T^* M$ exist mapping the boundary of the worldsheet Riemann surface $\Sigma$ to $M$. Nevertheless, the A-model on this geometry is not trivial, as the partition function receives contributions from the boundary of the moduli space of Riemann surfaces, where these degenerate to graphs. Witten argues, using open string field theory, that this contribution is captured by the partition function of Chern-Simons theory on $M$. The Chern-Simons theory depends on the level $k$ and the rank $N$ of the gauge group. In the open string description, $N$ maps to the number of branes wrapping $M$, and the string coupling constant is given by
\be
g_s = \frac{2\pi}{k+N} \,.
\ee
The combination on the right-hand side of the above equation of level and rank is the familiar quantum correction to the naive coupling constant $\frac{1}{k}$ of Chern-Simons theory. 

A general Calabi-Yau manifold with Lagrangian submanifold $M$ will generically exhibit holomorphic curves ending on $M$. Witten instructs us to add a Wilson loop contribution
\be  \label{wilson_loop_cont}
\tr \,\mathrm{P} \exp \int_C A
\ee 
to the Chern-Simons action for each such holomorphic curve, with $C$ the knot, i.e. a one real dimensional closed curve in the 3-manifold $M$, formed by the intersection of the curve with $M$. Unless noted otherwise, all traces are to be understood in the fundamental representation. Determining the set of such holomorphic curves is a difficult problem. They will generically occur in families, requiring us to make sense of a contribution to the Chern-Simons action involving an integral over a family of knots $C_\alpha$ with integrand (\ref{wilson_loop_cont}). This prescription is hence calculationally not feasible for an arbitrary geometry. With an assiduous choice of geometry, however, it represents the starting point for computing $\Ztop$ on arbitrary toric Calabi-Yau manifolds.

\subsection{Determining all holomorphic curves in $T^2$  fibrations}  \label{T2}
\begin{figure}[h]
\centering
\includegraphics[width=9cm]{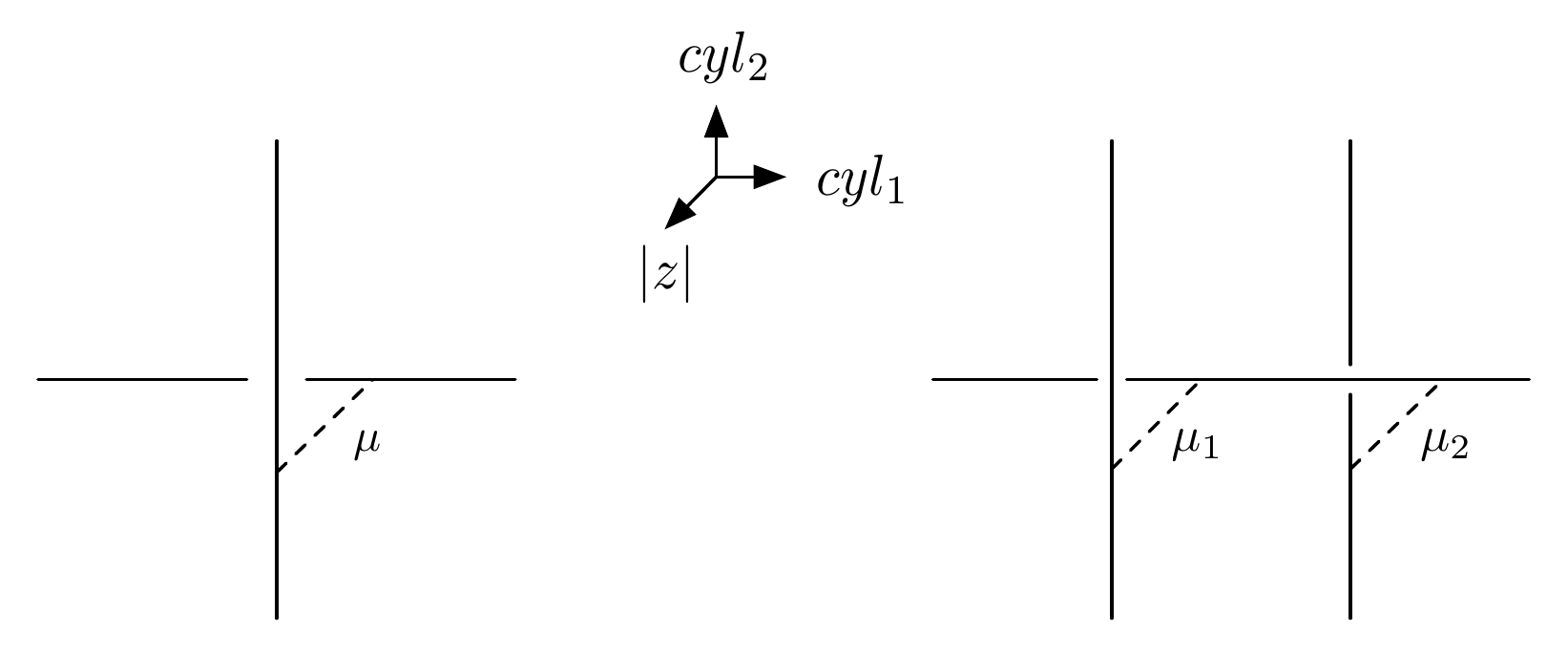}
\caption{Discriminant loci of $T^2$ fibrations.}  \label{fig:t2fib}
\end{figure}

The class of geometries we will consider \cite{Aganagic:2002qg} permit a $T^2$ torus action. The locus at which this action degenerates encodes the essential features of the geometry. The basic local building block of these geometries is the cotangent space of the 3-sphere, $T^* S^3$. This space carries a complex structure, inherited from $\IC^4$ via the embedding
\be  \label{cotans3}
xy = z \,, \quad uv = z + \mu  \,,
\ee
with $(x,y,u,v) \in \IC^4$, and $\mu \in \IC$ the complex structure modulus of the geometry. For $\mu$ real, the base $S^3$ of the cotangent space is situated at $x = \bar{y}$, $u = -\bar{v}$. The $T^2= S^1 \times S^1$ action on the geometry is given by $x \rightarrow e^{i \alpha}, y \rightarrow e^{-i \alpha}$ and $u \rightarrow e^{i \beta}, v \rightarrow e^{-i \beta}$. This action can be intuitively thought of as translation around the circle directions of the two cylinders defined by each of the two equations (\ref{cotans3}) individually at fixed $z$. One or the other $S^1$ action degenerates at $x=y=0$, $u=v=0$ respectively. Diagrammatically, the geometry can be depicted as in the left diagram in figure \ref{fig:t2fib}, in which the degeneration loci are represented in a real three dimensional slice of the geometry spanned by the axes direction of the two cylinders and $|z|$. The $S^3$ can be located along the dashed line (one quickly convinces oneself that a $T^2$ fibered over an interval such that one cycle degenerates at one end, a different cycle at the other end, is topologically an $S^3$). 

To obtain a geometry containing a holomorphic curve, we can modify these equations such that one of the torus cycles degenerates at two point of the $z$-plane:
\be  \label{b2=1}
xy = z \,, \quad uv = (z + \mu_1)(z-\mu_2)  \,.
\ee
The corresponding diagram is given on the right in figure \ref{fig:t2fib} for $\mu_1, \mu_2 > 0$. The geometry now contains an isolated holomorphic cylinder, given by $uv = - \mu_1 \mu_2$, which intersects each of the two 3-spheres indicated by dashed lines in an unknot.
Following Witten's prescription, the target space description of open topological string theory on this geometry is given by a Chern-Simons theory living on each $S^3$, with the holomorphic cylinder and its multicovers extending between the two 3-spheres giving rise to a Wilson loop insertion
\be  \label{cyl_contr}
\cO(U,V;r) = \exp \sum_{n=1}^\infty \frac{e^{-nr}}{n} \tr\, U^n \tr\, V^{-n} \,, \quad   U = \mathrm{P}  \exp \oint_{\gamma_1} A_1 \,,\quad V = \mathrm{P} \exp \oint_{\gamma_2} A_2 \,,
\ee
in the Chern-Simons path integral. The integration contours $\gamma_i$ are the unknots along which the cylinder intersects the two 3-spheres, and the coordinate $r$ is an open string modulus indicating the length of the cylinder, complexified by the classical vacuum expectation value of the Wilson loops. The two Chern-Simons theories can be essentially decoupled by an application of the Frobenius formula \cite{Labastida:2000zp,Aganagic:2002qg}, which yields the identity
\be  
\cO(U,V;r) = \sum_R e^{- l_R r} \tr_R U \,\tr_R V^{-1} \,.
\ee
The sum here is taken over all irreducible representations $R$ of $U(N)$, or equivalently over all possible Young diagrams with at most $N$ rows. We will drop the constraint on the number of rows by considering the large $N$ limit. $l_R$ indicates the first Casimir of the representation $R$, or equivalently the number of boxes of the corresponding Young diagram. Witten's prescription thus gives rise to the following partition function for the open string on the geometry on the right in figure \ref{fig:t2fib}:
\be  \label{part_cyl}
\Zop =  \sum_R e^{- l_R r} \langle \tr_R U \rangle_{1} \langle \tr_R V^{-1} \rangle_2  \,.
\ee
Computations on more intricate $T^2$ fibrations follow the same pattern. They will generically involve multiple cylinders ending on the same Lagrangian $S^3$ manifold. Each cylinder will intersect the $S^3$ in an unknot. Computing the corresponding Wilson loop factor requires determining the linking number of all unknots on a given $S^3$. In the following, we will only need the result for a Hopf link, the link of two unknots with linking number one. This was computed in \cite{MR1983094}, as a function of the Chern-Simons coupling $\log q = \frac{ 2 \pi i}{k+N}$ and the 't Hooft coupling $\lambda = q^N$. For reasons we will explain below, we will here be interested only in the leading $\lambda$ limit of this result, which we denote as ${\cal W}_{R_{1}R_{2}}(q)$ for the case in which the two linked unknots carry representations $R_1$, $R_2$ respectively. Explicit expressions can be found in \cite{MR1983094, Aganagic:2002qg}. Organizing the calculation in terms of the size of representations $l_R$, one can thus compute $\Zop$ to arbitrary order in the open string modulus $e^{-r}$.

\newpage

\subsection{From open to closed via geometric transitions}

\subsubsection{The conifold transition}
\begin{figure}[h]
\centering
\includegraphics[width=9cm]{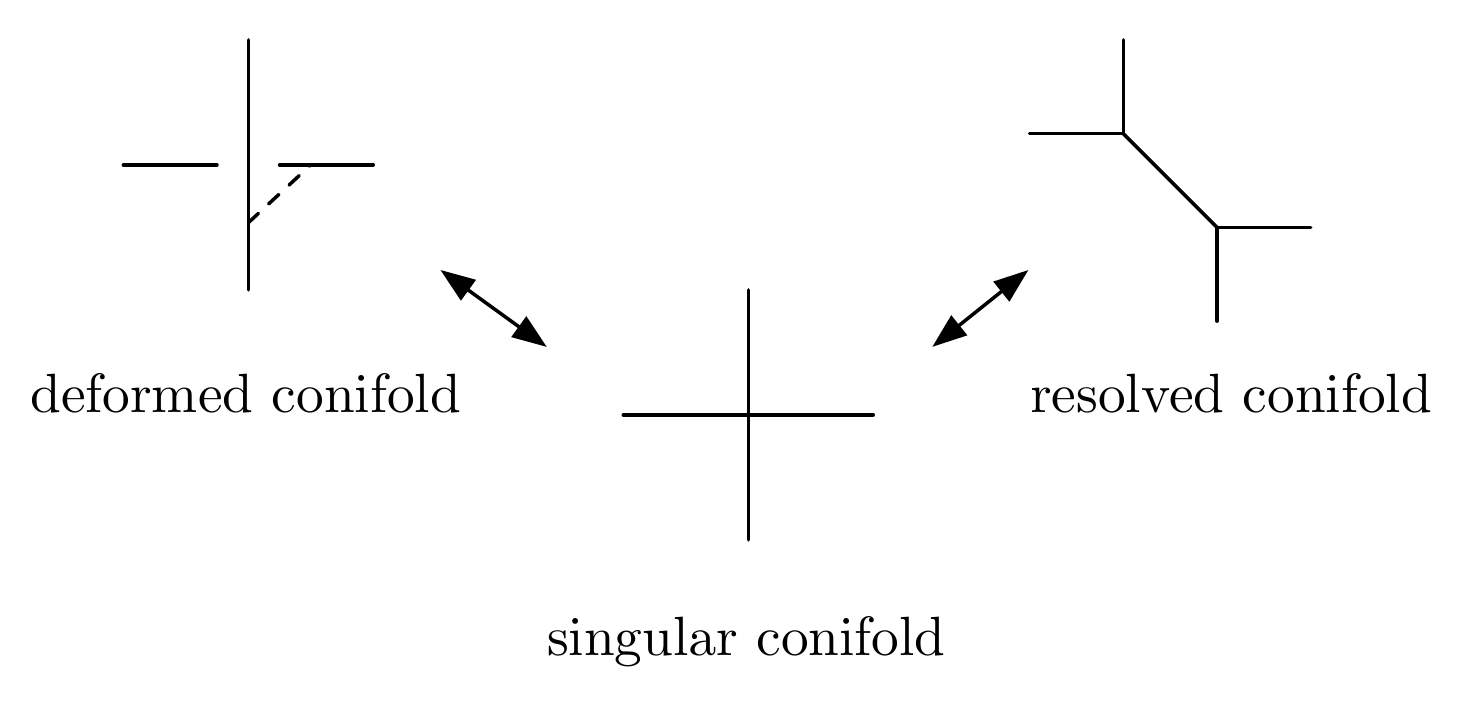}
\caption{The conifold transition diagrammatically.}  \label{fig:conifold_trans}
\end{figure}

In \cite{Gopakumar:1998vy}, Gopakumar and Vafa proposed an open-closed duality for the topological string, similar in spirit to the AdS/CFT correspondence of the full physical string theory: the theory in the presence of branes is dual to a theory without branes on a modified geometry. The simplest example of this duality relates the open topological A-model on $T^* S^3$ with $N$ branes wrapping the Lagrangian $S^3$ at the base of the fibration, which we encountered above, to the closed topological A-model on the sum of two line bundles over the projective line, $\cO(-1) \oplus \cO(-1) \rightarrow \IP^1$. The complexified K\"ahler parameter $t$ of the $\IP^1$ base of this geometry is related to the number of branes wrapping the submanifold $S^3$ of $T^*S^3$ via
\be  \label{open_to_closed}
t = i\,g_s N  \,.
\ee
The geometries on the two sides of the duality are related to each other as follows. Deforming the complex structure of $T^* S^3$ all the way to $\mu = 0$ gives rise to a singular manifold, called the conifold, defined by the equation
\be
xy - uv = 0
\ee
in $\IC^4$. This space is a local model for a common singularity in Calabi-Yau manifolds. To smoothen the singularity, one can deform the space by introducing the complex parameter $\mu$ as in (\ref{cotans3}), to re-obtain $T^* S^3$. The cotangent space of $S^3$ is therefore also referred to as the deformed conifold. A second possibility is to resolve the singularity via a general procedure in algebraic geometry referred to as a blow-up, in which a point is replaced by a projective space. This yields the space $\cO(-1) \oplus \cO(-1) \rightarrow \IP^1$, which is therefore also called the resolved conifold. The size of the $\IP^1$ is measured by the K\"ahler parameter $t$, and the singular limit is reached at $t \rightarrow 0$. The passage from the deformed to the resolved conifold is termed the conifold transition. Diagrammatically, it can be represented as in figure \ref{fig:conifold_trans}. 

\newpage

\subsubsection{Toric geometry and web diagrams} \label{s:web}
Note that at the point $\mu = 0$ of (\ref{cotans3}), a third $S^1$ action can be defined on the geometry: multiplication of $x$ and $u$ by a common phase. A $T^3$ action in three complex dimensions is the hallmark of a three complex dimensional toric manifold. The toric property is preserved by the blow-up which resolves the singularity. The diagram in figure \ref{fig:conifold_trans} representing the resolved conifold is called a web diagram. Its distinguishing features are the trivalent vertex, rational edges, and the fact that the sum of the primitive vectors emanating from each vertex vanishes. Any such diagram encodes a toric Calabi-Yau geometry. These diagrams will play a central role in the formulation of the topological vertex, to which we shall turn in subsection \ref{s:vertex}. 

\begin{figure}[h]
\centering
\includegraphics[width=9cm]{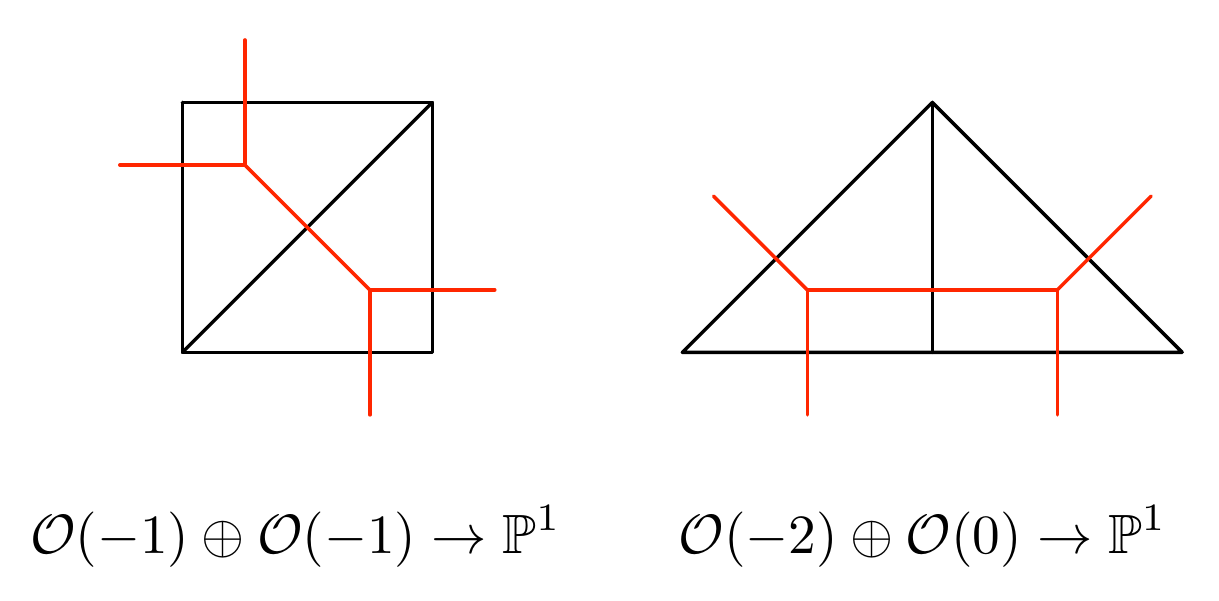}
\caption{Toric fans and their dual web diagrams.}  \label{fig:dual_web_fan}
\end{figure}

A more general formalism for describing toric geometry, valid beyond the context of local Calabi-Yau manifolds, proceeds by encoding the geometry in terms of toric fans (see e.g. the classic exposition in \cite{fulton}). In the case of local Calabi-Yau manifolds, these diagrams are dual to web diagrams, in the sense depicted in figure \ref{fig:dual_web_fan}.

\subsubsection{Generalized conifold transitions}
The authors of \cite{Diaconescu:2002sf,Aganagic:2002qg} proposed extending the conifold transition to the $T^2$ fibrations encountered in subsection \ref{T2}, by performing such transitions locally on each $S^3$. Conversely, given a toric Calabi-Yau manifold, one can introduce additional $\IP^1$ cycles, called exceptional cycles, via the blow-up procedure alluded to above. Provided that the resulting manifold retains the Calabi-Yau property, the local geometry of the blow-up cycle will be that of a resolved conifold. Performing the conifold transition on a sufficient number of such cycles should give rise to the class of geometries discussed above. Naturally, care must be exercised that these local manipulations be consistent with the global constraints of the geometry. Under these manipulations, holomorphic curves intersecting two exceptional cycles map to holomorphic cylinders intersecting the 3-spheres the cycles transition into. The corresponding K\"ahler parameters map to open string moduli.

\begin{figure}[h]
\centering
\includegraphics[width=11cm]{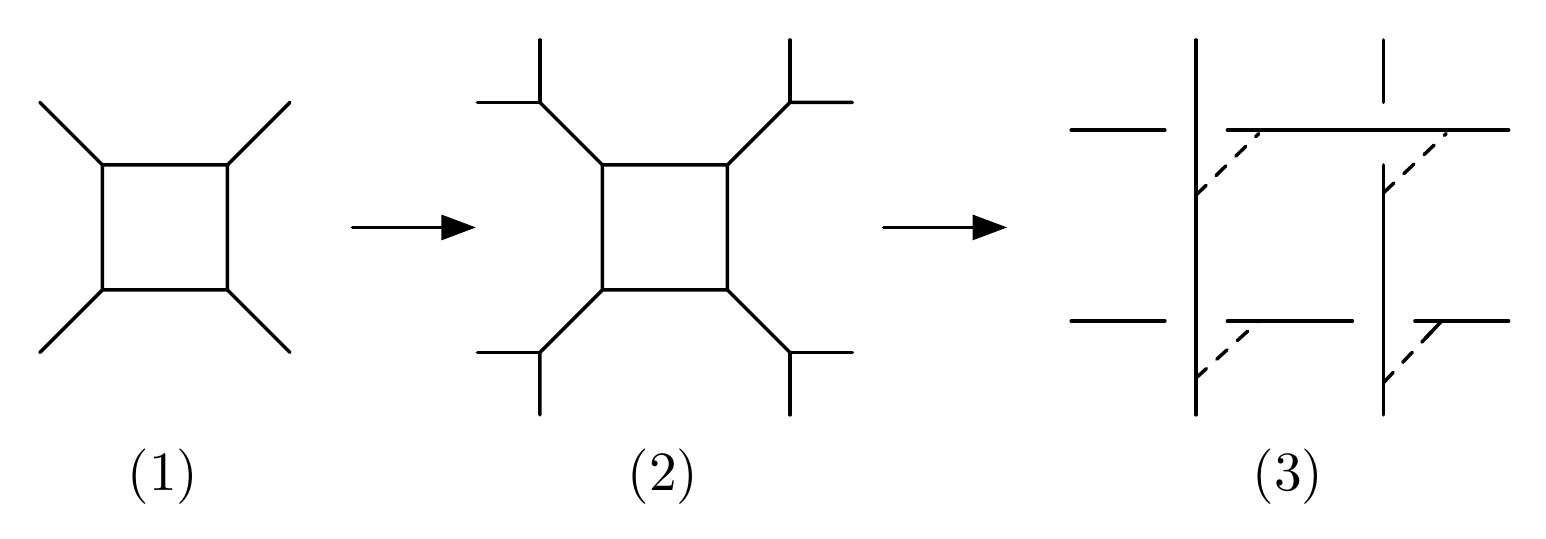}
\caption{Manipulations leading from local $\IP^1 \times \IP^1$ to an open geometry.}  \label{fig:p1p1_trans}
\end{figure}

We have sketched these manipulations for the case of local $\IP^1 \times \IP^1$,\footnote{Given a complex surface $M$, the total space of its canonical line bundle is a non-compact Calabi-Yau manifold, referred to as `local $M$'. Non-compact Calabi-Yau manifolds are also referred to as local Calabi-Yau, the nomenclature indicating that they should be thought of as the local description of a larger, compact geometry.} which will serve as a principal example in the next subsection, in figure \ref{fig:p1p1_trans}. 
The diagram (1) is the web digram representing local $\IP^1 \times \IP^1$. Blowing up the four toric fixed points, we obtain diagram (2). Performing local conifold transitions on the four exceptional cycles then yields diagram (3). The partition function of the open $A$-model on this latter geometry can be computed via Chern-Simons theory as described above. This coincides with the closed topological string partition function of geometry (2) upon the correct mapping of open to closed variables. To recover the original geometry, we can take the $t \rightarrow \infty$ limit of all K\"ahler parameters associated to exceptional curves. As is evident from the general form of the partition function (\ref{ztop_gv}), all contributions involving exceptional curves are thus set to 0. By the map (\ref{open_to_closed}) between closed and open variables, the limit in terms of open variables is given by $\lambda \rightarrow \infty$.\footnote{In fact, as first observed in \cite{Diaconescu:2002sf}, an additive correction term is required to the naive identification between open and closed moduli when geometric transitions beyond the simple case of the conifold are considered. A first principles derivation of this shift in the context of a proof of the open/closed duality of geometric transitions, e.g. along the lines of \cite{Ooguri:2002gx}, is still outstanding.} This is the rationale behind the introduction above of $\cW_{R_1 R_2}(q)$ as the $\lambda \rightarrow \infty$ limit of the Hopf link expectation value.

%Due to the leading $\lambda$ dependence of $W_{R_1,R_2}$ given in (\ref{link}), this cannot be quite right. One can easily cancel this $\lambda$ dependence of the partition function by defining 
% \be  \label{shift}
% T_B = r_B + \frac{t_2 + t_3}{2} =  r_B + \frac{t_1 + t_4}{2}  \,, \\
% T_F = r_F + \frac{t_1 + t_2}{2} =  r_F + \frac{t_3 + t_4}{2} \,.
% \ee
% 

\subsection{Computing: local $\IP^1 \times \IP^1$}  \label{s:local_p1p1}
\subsubsection{Performing sums over representations}
We have reduced the computation of the partition function of the closed topological string on local $\IP^1 \times \IP^1$ to the evaluation of the following expression:
\be  \label{zp1p1}
\Ztop =\sum_{R_{1,2,3,4}}Q_B^{-(l_{1}+l_{3})}
Q_F^{-(l_{2}+l_{4})} {\cal W}_{R_{1}R_{4}}(q)\,{\cal
  W}_{R_{4}R_{3}}(q) {\cal W}_{R_{3}R_{2}}(q){\cal W}_{R_{2}R_{1}}(q) \,,
\ee
where the Hopf link expectation values $\cW_{R_i R_j}$ were defined at the end of subsection \ref{T2}, and $Q_{B,F}=e^{-t_{B,F}}$ denote the exponentiated K\"ahler parameters associated to the two $\IP^1$-cycles. By the results of \cite{MR1983094} for the computation of Wilson loop expectation values involving the Hopf link, we can evaluate this expression to arbitrary order in $Q_B$ and $Q_F$. These calculations rapidly become cumbersome. In \cite{Iqbal:2003ix}, Iqbal and I demonstrated how to sum the series over $Q_F$ exactly, with implications involving an important conjecture by Nekrasov, as we discuss in subsection \ref{rep_nek}. We will indicate the scope of these summation techniques in subsection \ref{s:vertex}, based on \cite{Iqbal:2004ne}. For now, continuing with the example at hand, let us introduce
\be  \label{def_K}
K_{R_{1}R_{2}}(Q)=\sum_{R}Q^{l_{R}}{\cal W}_{R_{1}R}(q)\,{\cal W}_{RR_{2}}(q)
\ee
as a building block of the partition function. The crucial step in the summation is writing
\be 
K_{R_{1}R_{2}}(Q)=
K_{\cdot \cdot}(Q){\cal W}_{R_{1}}(q){\cal W}_{R_{2}}(q) \exp\left[ \sum_{n=1}^{\infty}f_{R_{1}R_{2}}^{n}(q)Q^{n}\right] \,,
\ee
where
\be  \label{-20curve}
 K_{\cdot \cdot}(Q) = \exp \left[\sum_{n=1}^\infty f^n_0(q) \right] \,, \quad f^n_0(q) = \frac{f^1_0(q^n)}{n} \,, \quad f^1_0(q) = \frac{q}{(q-1)} 
\ee
is the partition function for a (-2,0)-curve \cite{Aganagic:2002qg}, and making the ansatz
\be
f_{R_{1}R_{2}}^{n}(q)=\frac{f_{R_{1}R_{2}}(q^{n})}{n}\,.
\ee
This ansatz, mirroring (\ref{-20curve}), reflects the fact that the contributions to $K_{R_1 R_2}$ stem from a single isolated curve and its multicovers. It was proved rigorously in \cite{Eguchi:2003sj}. We can determine $f_{R_{1}R_{2}}(q)$ by computing (\ref{def_K}) to first order in $Q$. We obtain
\ba
f_{R_{1}R_{2}}(q)&=&\frac{{\cal W}_{R_{1} \tableau{1}}}{{\cal W}_{R_{1}}}
\frac{{\cal W}_{\tableau{1} R_{2}}}{{\cal W}_{R_{2}}}-{\cal W}_{\tableau{1}}^{2}\,\\
&=&
\frac{q}{(q-1)^{2}} \left[ \left( 1+(q-1)\sum_{j=1}^{d_{1}}(q^{\mu^1_{j}-j}-q^{-j}) \right)
\left( 1+(q-1)\sum_{j=1}^{d_{2}}(q^{\mu^2_{j}-j}-q^{-j})  \right) -  1 \right] \,\nn.
\ea
We have here denoted the Young diagram corresponding to representation $R_i$ as $\mu^i$, with $d_i$ indicating the number of rows and $\mu^i_j$ the number of boxes in its $j$-th row. This expression can be further simplified to
\ba \label{frr}
f_{R_{1}R_{2}}(q)&=&(q-2+q^{-1})f_{R_{1}}(q)f_{R_{2}}(q)+f_{R_{1}}(q)+f_{R_{2}}(q)\nn\\
&=:&
\sum_{k}C_{k}(R_{1},R_{2})q^{k}\,,
\ea
with
\be \label{fr}
f_{R}(q):=f_{R,\,.}(q) = \sum_{j=1}^{d}\sum_{v=1}^{\mu_{j}}q^{v-j}\,.
\ee
The partition function (\ref{zp1p1}) thus takes the form
\be
\Ztop=K^{2}_{\cdot \cdot}(Q_{F})\sum_{R_{1},R_{2}}Q_{B}^{l_{R_{1}}+l_{R_{2}}}\frac{{\cal W}^{2}_{R_{1}}(q){\cal W}^{2}_{R_{2}}(q)}
{\prod_{k}(1-q^{k}Q_{F})^{2C_{k}(R_{1},R_{2})}}\,.
\label{pf}
\ee

\subsubsection{Computing Gopakumar-Vafa invariants}
Extracting Gopakumar-Vafa invariants from (\ref{pf}) is an easy exercise. Since we have summed the series over $Q_F$, we introduce the following parametrization of the general Gopakumar-Vafa form (\ref{ztop_gv}) of the partition function:
\be
\mathrm{Z_{top}^{GV}} = K_{\cdot\,\cdot}(Q_{F})^{-n^{0}_{(0,1)}}\,\exp \{\sum_{n=1}^{\infty}\frac{1}{n}\sum_{k=1}^{\infty}Q_{B}^{kn}G_{k}(q^{n},Q_{F}^{n})\} \,,
\ee
where
\be
G_{k}(q,Q_{F})=\sum_{m=0}^{\infty}\sum_{g=0}^{\infty}
\frac{{n}^{g}_{(k,m)}\,(-q)^{1-g}}{(q-1)^{2-2g}}\,Q_{F}^{m}=\sum_{g=0}^{\infty}\frac{1}{(q^{1/2}-q^{-1/2})^{2-2g}}\,f_{g}^{(k)}(Q_{F})\,.
\ee
To extract the factor $K_{\cdot\,\cdot}(Q_{F})$, we have invoked the fact that $n^g_{(0,m)} \sim \delta_{g,0} \delta_{m,1}$ \cite{KKV}. We further parametrize the partition function (\ref{pf}) as 
\be
\Ztop =K_{\cdot \,\cdot}^{2}(Q_{F})\,\sum_{k=0}^{\infty}Q_{B}^{k}Z_{k}(Q_{F},q)\,,
\ee
with
\be
Z_{k}(q,Q_{F})=\sum_{\{R_1,R_2|l_{R_1}+l_{R_2}=k\}}\frac{{\cal W}_{R_{1}}^{2}\,
{\cal W}_{R_{2}}^{2}}{\prod_{m}(1-q^{m}Q_{F})^{2C_{m}(R_{1},R_{2})}}\,.
\ee
Equating $\mathrm{Z_{top}^{GV}}=\Ztop$, we deduce ${n}^{0}_{(0,1)}=-2$ (a result obtained in \cite{KKV}) and 
\ba \label{relations}
G_{1}(q,Q_{F}) &=& Z_{1}(q,Q_{F})\,,\\ \nn
G_{2}(q,Q_{F}) &=&Z_{2}(q,Q_{F})- \frac{1}{2}Z_{1}(q,Q_{F})^{2}-\frac{1}{2}Z_{1}(q^2,Q_{F}^2)\,, \\ \nn
&etc.&
\ea
Evaluating (\ref{pf}) explicitly, we find for $k= 1, \ldots, 4$
\be  \label{fgtp}
f^{(k)}_{g}(x)=\frac{P^{(k)}_{g}(x)}{(1-x)^{2g+4k-2}}\,,
\ee
where the functions $P^{(k)}_{g}(x)$ are finite at $x=1$. 
%This behavior will become important when considering the field theory limit in the next subsection. Explicitly,
To give an impression of the expressions involved, we cite the first few results from \cite{Iqbal:2003ix}:
\be
f^{(1)}_{0}(x)=-\frac{2}{(1-x)^{2}}\,,\,\,\,\,\,f^{(1)}_{g>0}(x)=0\,,
\ee
\be
f^{(2)}_{g}(x)=\frac{(3g+6)x^{g+1}+(6g+8)x^{g+2}+(3g+6)x^{g+3}}{(1-x)^{2g+6}(1+x)^{2}}\,.
\ee
%Note that the result for $k=1$ can be obtained directly from the geometry: since all curves $B+mF$ are of genus zero, the moduli space of curves is just $\IP^{2m+1}$ and therefore $N^{g}_{(1,m)}=-(2m+2)\delta_{g,0}$, consistent with $Z_{1}(q,Q_{F})$.

\subsection{Reproducing Nekrasov's partition function via geometric engineering}  \label{rep_nek}
\subsubsection{Nekrasov's partition function}
In \cite{Nekrasov:2002qd}, Nekrasov computed the prepotential of $\cN=2$ gauge theories, previously obtained by Seiberg and Witten by what is now known as Seiberg-Witten theory (we will have much to say about this in section \ref{section:hae}), directly by computing integrals on instanton moduli space via localization methods. The idea of applying localization to compute integrals over instanton moduli space was already presented in \cite{Losev:1997tp,Losev:1997wp,Fucito:2001ha,Dorey:2000zq,Hollowood:2002ds}. Nekrasov demonstrated that enlarging the equivariant action on moduli space by the two commuting (Euclidean) spacetime rotations of orthogonal planes leads to isolated fixed points, and computed their contribution to the relevant integrals as a function of the two additional equivariant parameters, $\epsilon_1$ and $\epsilon_2$. Assembling these results in a generating function weighted by instanton number gives rise to the Nekrasov partition function
\be
\Znek(\epsilon_1,\epsilon_2)  = \sum_n \mathfrak{q}^n \int_{\tilde{M}_n} \omega  \,,
\ee
where we have denoted the relevant $n$-instanton moduli space as $\tilde{M}_n$ and $\omega$ stands for an appropriate equivariant form depending on the details of the gauge theory under consideration. The instanton counting parameter $\mathfrak{q}$ is related to the ultraviolet gauge coupling. $\Znek$ contains much information beyond the prepotential of the four dimensional gauge theory, the computation of which was the primary motivation underlying \cite{Nekrasov:2002qd}. Nekrasov conjectured that it coincides with a counting function capturing BPS states in the spectrum of the gauge theory embedded within string theory. At $\epsilon_1=-\epsilon_2$, the counting function specializes to the index underlying (\ref{ztop_gv}), $\Znek$ hence, according to this conjecture, to $\Ztop$. Proving this conjecture was part of the motivation behind the works \cite{Iqbal:2004ne,Iqbal:2003zz,Iqbal:2003ix}. 

For the case of pure $SU(2)$ gauge theory, the Nekrasov partition function at $\epsilon= \epsilon_1= - \epsilon_2$ takes the form
\be
\Znek=\sum_{\mu^1,\mu^2} \mathfrak{q}^{|\mu^{1}|+|\mu^{2}|}\,\prod_{l,n=1,2}\prod_{i,j=1}^{\infty}\frac{\sinh \frac{\beta}{2}(a_{ln}+\epsilon(\mu^l_{i}-\mu^n_{j}+j-i))}{\sinh \frac{\beta}{2} (a_{ln}+\epsilon(j-i))}\,.  \label{z_nek}
\ee
The sum is over all fixed points of the equivariant action, which are in one-to-one correspondence with Young diagrams $\mu^1$, $\mu^2$. The number of boxes of a Young diagram $\mu$ is indicated by $|\mu|$. $a_{12}=-a_{21}=2a$ is the vacuum expectation value of the adjoint scalar (see section \ref{s:sw_rev}), $a_{11}=a_{22}=0$. Factors with $(l,i) = (n,j)$ are understood to be equal to 1. The parameter $\beta$ is best thought of as the radius of an additional circle which lifts the four dimensional gauge theory to a five dimensional one \cite{Lawrence:1997jr}. The four dimensional result is obtained in the $\beta \rightarrow 0$ limit.

\subsubsection{$\Znek$ from topological string theory} \label{s:geom_eng}

Local $\IP^1 \times \IP^1$, the geometry studied in subsection \ref{s:local_p1p1}, is the simplest example of a Calabi-Yau geometry whose effective field theory description exhibits enhanced gauge symmetry. Using e.g. toric methods, it is not difficult to see that this geometry coincides with the resolution of an $A_1$ singularity fibered over $\IP^1$. The singular geometry yields enhanced gauge symmetry, as D2 branes, which give rise to W-bosons in the gauge theory, become massless here. Embedding a gauge theory into string theory by an appropriate choice of geometry is referred to as {\it geometric engineering}. The mapping of parameters between geometry and gauge theory in the case of local $\IP^1 \times \IP^1$ and pure $SU(2)$ gauge theory is the following: 
\be
\frac{Q_B Q_F}{2^4}=\mathfrak{q}  \,, \quad Q_F = \exp 2 \beta a  \,, \quad  g_s^2 = \beta^2 \epsilon_1 \epsilon_2 \,.
\ee
$Q_F$ here encodes the size of the cycle resolving the $A_1$ singularity. It consequently maps to the vacuum expectation value $a$ of the adjoint scalar of the gauge theory, which induces the breaking of the gauge symmetry. The ultraviolet gauge coupling is inversely proportional to the size of the compactification manifold.

To relate (\ref{pf}) to (\ref{z_nek}), we note the equality \cite{Iqbal:2003ix}
\be
{\cal W}_{R}^{2}(q)=2^{-2l_{R}}\,q^{\kappa_{R}/2}\prod_{i,j=1}^{\infty}\frac{\sinh \frac{\beta}{2}\epsilon(\mu_{i}-\mu_{j}+j-i)}{\sinh \frac{\beta}{2} \epsilon(j-i)} \,.
\ee
Hence, the ${\cal W}_R$ terms in (\ref{pf}) account for the factors in the product (\ref{z_nek}) with $l=n$. The remaining terms follow from the identity
\ba 
\prod_{k}(1-q^{k}Q_{F})^{-2C_{k}(R_{1},R_{2}^{T})}&=& Q_{F}^{-l_{R_{1}}-l_{R_{2}}}2^{-2(l_{R_{1}}+l_{R_{2}})}q^{-\frac{1}{2}(\kappa_{R_{1}}-\kappa_{R_{2}})}\\ 
&&\prod_{l\neq n,i,j}\frac{\sinh \frac{\beta}{2}(a_{ln}+\epsilon(\mu^l_{i}-\mu^n_{j}+j-i))}{\sinh \frac{\beta}{2}(a_{ln}+\epsilon(j-i))} \,.\nn
\ea
This identity was proposed and checked experimentally in \cite{Iqbal:2003ix}, and proved in \cite{Eguchi:2003sj}. 

It is possible to use a generalization of the methods described above to perform computations on local Calabi-Yau manifolds resolving higher $A_n$ singularities, related to gauge theories of higher rank \cite{Iqbal:2003zz}. These computations however become cumbersome, and are best performed within the framework of the topological vertex \cite{Aganagic:2003db}.

\subsubsection{Beyond $\epsilon_1 = - \epsilon_2$} \label{s:beyond_gae}
The physical meaning of $\Znek$ beyond the $\epsilon_1 = - \epsilon_2$ locus has been elucidated from several perspectives in the decade since the publication of \cite{Nekrasov:2002qd}. In \cite{Iqbal:2007ii}, generalizing the computation that led to (\ref{ztop_gv}), $\Znek$ at general $(\epsilon_1$, $\epsilon_2)$ was reformulated as a counting function of BPS states, thus extending its definition beyond the gauge theory context. The same reference also introduces a refinement of the topological vertex (we will introduce the unrefined vertex below) to compute this counting function, christened the refined topological string partition function, for arbitrary toric geometries. Refined topological string amplitudes $F^{(n,g)}$ are formally defined as the expansion coefficients of this partition function in $g_s^2 = \epsilon_1 \epsilon_2$ and $s = (\epsilon_1 + \epsilon_2)^2$,
\be
\Ztop(g_s,s) = \exp \sum_{g,n} F^{(n,g)} g_s^{2g-2} s^n \,.
\ee
In \cite{Nekrasov:2009rc}, a Lagrangian was formulated for a gauge theory with partition function $\Znek(\epsilon_1, \epsilon_2)$, referred to as the $\Omega$-deformation of the initial gauge theory, and a relation to two dimensional integrable systems established. The worldsheet definition of this refinement is still not completely settled. Steps in this direction have been taken e.g. in \cite{Antoniadis:2013bja}, where the role of the refined amplitudes in $\cN=2$ supergravity is investigated. Mathematically, refinement has been interpreted in terms of motivic invariants \cite{Dimofte:2009bv,Choi:2012jz}. A remarkable correspondence between $\Omega$-deformed gauge theory and two dimensional conformal field theory was proposed in \cite{AGT}. This will play a central role in section~\ref{s:agt}.

\subsection{The topological vertex on the strip} \label{s:vertex}
\subsubsection{The general formalism}
The above calculation already suggests the emergence of diagrammatic rules for the computation of $\Ztop$ for local toric Calabi-Yau geometries. The principal complication in the direct computation of the partition function with the methods described is that the web diagrams underlying toric Calabi-Yau manifolds, introduced in subsection \ref{s:web}, exhibit a trivalent vertex, while the manipulations exemplified in figure \ref{fig:p1p1_trans} give rise to a tetravalent vertex as building block, and require taking limits to access general toric geometries. In \cite{Aganagic:2003db}, an expression for a trivalent vertex was extrapolated from the circle of ideas presented here. In the form presented in \cite{Okounkov:2003sp}, it is given by
\begin{eqnarray}  \label{vertex}
C_{\lambda \mu \nu} = q^{\frac{\kappa(\lambda)}{2}}s_{\nu}(q^\rho)\sum_\eta s_{\lambda^t/\eta}(q^{\nu+\rho})s_{\mu/\eta}(q^{\nu^t+\rho}) \,.
\end{eqnarray}
The indices $\lambda, \mu, \nu$ symbolize Young diagrams. A sum over such an index indicates a sum over all possible Young diagrams. $s_\mu$ denotes the Schur function associated to the diagram $\mu$. Schur functions form a particular basis for symmetric polynomials, with the number of variables here taken to infinity. The notation is $s(q^{\nu+\rho}) = s(\{q^{\nu_i -i +\frac{1}{2}}\})$, with $\nu_i$ indicating the number of boxes in the $i$-th row of the diagram $\nu$. The $s_{\mu / \eta}$ are skew Schur functions, defined by
\ba \label{skew}
s_{\mu / \eta} = \sum_\nu c^{\mu}_{\eta \nu} s_\nu  \,,
\ea
where the $c^{\mu}_{\eta \nu}$ are tensor product coefficients, and $\kappa(\lambda) = \sum \lambda_i(\lambda_i - 2i+1)$. Though not manifest, the expression (\ref{vertex}) enjoys cyclic symmetry with regard to the Young diagrams $\lambda, \mu, \nu$. Two vertices are glued, up to a subtlety to which we shall return momentarily, by performing a sum over Young diagrams as follows,
\be
\sum_\lambda C_{\lambda \mu \nu} C_{\lambda^t \rho \sigma} Q^{|\lambda|}  \,.
\ee
Here, $\lambda^t$ is the transposed Young diagram, obtained by exchanging rows and columns, and $|\lambda|$, as above, indicates the total number of boxes of the Young diagram $\lambda$. $Q$ is the exponentiated K\"ahler parameter corresponding to the cycle obtained by gluing the vertices. The subtlety in gluing vertices is related to the necessity of specifying a framing when computing Wilson loop vacuum expectation values in Chern-Simons theory (we glossed over the framing question in the discussion above, it is of course addressed in \cite{Iqbal:2003ix,Iqbal:2003zz}). In the vertex formalism, the framing information translates into an integral vector associated to each leg of the vertex. Changes of framing with regard to the canonical framing $f_i$ chosen in (\ref{vertex}) are encoded in a triplet of integers $n_i$ \cite{Aganagic:2003db}, 
\ba
C_{\alpha_1 \alpha_2 \alpha_3}^{f_1 - n_1 v_1,f_2 - n_2 v_2,f_3 - n_3 v_3} &=& (-1)^{\sum_i n_i |\alpha_i|} q^{\sum_i n_i \frac{\kappa_{\alpha_i}}{2}} C_{\alpha_1 \alpha_2 \alpha_3}^{f_1,f_2,f_3}  \,.
\ea 
The gluing rules require gluing vertices with opposite framing vectors. One can easily check that the particular choice of framing vectors, as long as the choices on the two glued legs are correlated and opposite, is irrelevant upon performing the sum over Young diagrams.

\subsubsection{Computing the partition function on the strip}
\begin{figure}[h]
\centering
\includegraphics[width=5cm]{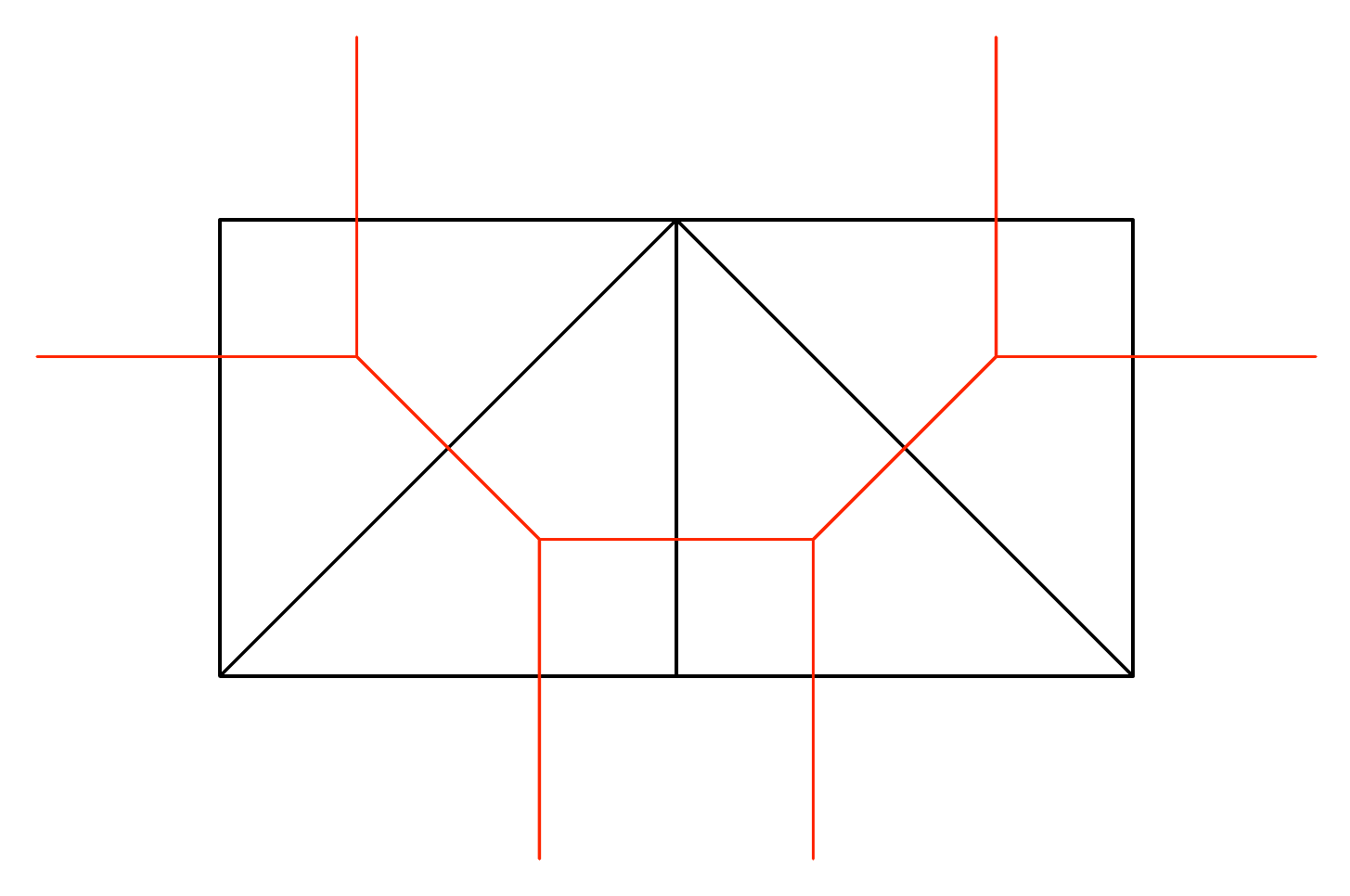}
\caption{A fan obtained by triangulating a strip, and the corresponding web diagram.}  \label{fig:strip}
\end{figure}

In \cite{Iqbal:2004ne}, Iqbal and I set out to study methods for performing the sums over Young diagrams explicitly in the context of the topological vertex. Based on the two identities \cite{MR1354144}
\ba  \label{sum_id}
\sum_\alpha s_{\alpha/\eta_1}(x)s_{\alpha/\eta_2}(y)&=&  \prod_{i,j} (1-x_i y_j)^{-1} \sum_\kappa s_{\eta_2/\kappa}(x)s_{\eta_1/\kappa}(y) \,, \\
\sum_\alpha s_{\alpha^t/\eta_1}(x)s_{\alpha/\eta_2}(y) &=&  \prod_{i,j} (1+x_i y_j) \sum_\kappa s_{\eta_2^t/\kappa^t}(x)s_{\eta_1^t/\kappa}(y) \,, \nn
\ea
we demonstrated that the sums occurring in the evaluation of the partition function on any triangulation of the strip, in the sense exemplified in figure \ref{fig:strip}, can be performed. 

These strips in turn can be used as building blocks for more intricate geometries, as we will discuss below. The geometry encoded by a strip diagram consists of two types of curves strung together: $\cO(-1) \oplus \cO(-1) \rightarrow \IP^1$, referred to as a (-1,-1) curve, and $\cO(-2) \oplus \cO \rightarrow \IP^1$, referred to as a (-2,0) curve. The corresponding fans and web diagrams are depicted in figure \ref{fig:dual_web_fan}. The contribution from such curves to $\Ztop$ can be easily computed, using the vertex and the summation identities (\ref{sum_id}). For a $(-1,-1)$ curve, it is
\be \label{c11}
\{ \alpha \beta \}_Q := \prod_k  (1- Q q^k)^{ C_k(\alpha, \beta)} \exp \left[ \sum_{n=1}^\infty \frac{Q^n}{n(2 \sin( \frac{n g_s}{2}))^2} \right] \,,
\ee
and for a $(-2,0)$ curve,
\be \label{c20}
[\alpha \beta]_Q :=\frac{1}{\{ \alpha \beta \}_Q  }  \,.
\ee
The exponential factor yields $\Ztop$ of these curves by themselves, i.e. not as building blocks of a larger geometry, as its $(\alpha,\beta)$-dependent coefficient is set to one by choosing these diagrams trivial, $\alpha = \beta = \cdot$. The Gopakumar-Vafa invariants of these simple geometries can be read off from these expressions as $n^g_m = \pm \delta_{g,0} \delta_{m,1}$. 

Using this notation, the link invariant $W_{\alpha \beta}$, which is the tetravalent vertex on which our calculation in the previous subsections was based (with $\cW_{\alpha \beta}$ as its $\lambda \rightarrow \infty$ limit), is related to the topological vertex as follows \cite{Iqbal:2004ne}:
\be
W_{\alpha \beta}=  \lambda^{\frac{|\alpha|+|\beta|}{2}} \sum_\gamma\frac{ C_{\beta^t \gamma \alpha} C_{\bullet \bullet \gamma^t} (-1)^{|\gamma|} Q^{|\gamma|}}{\{\bullet \bullet\}_{Q}}   \,, 
\ee
where $Q= \lambda^{-1}$. 

\begin{figure}[h]
\centering
\includegraphics[width=4cm]{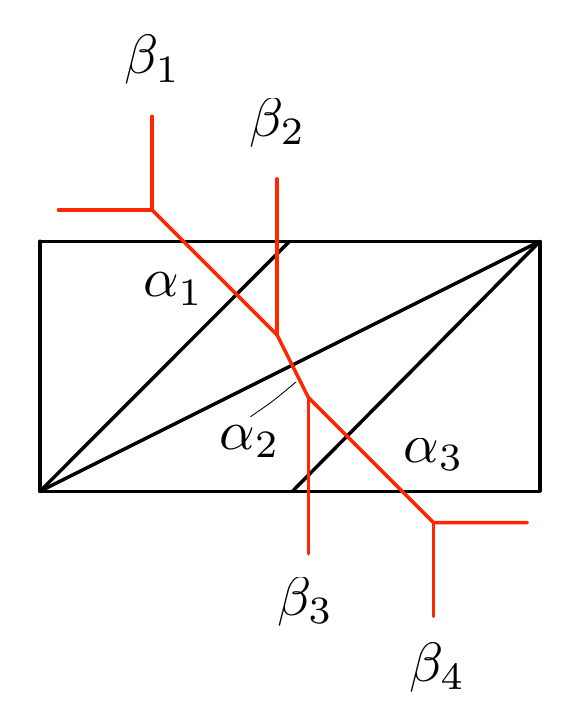}
\caption{Labelled web diagram corresponding to (\ref{strip_example}).}  \label{fig:strip_example}
\end{figure}

The partition function on a triangulated strip consists of a product of contributions (\ref{c11}) and (\ref{c20}), one for each pairing of vertices, with the corresponding K\"ahler parameter $\log \,Q$ being the sum of all intermediate K\"ahler parameters \cite{Iqbal:2004ne}. E.g., the diagram depicted in figure \ref{fig:strip_example} yields the partition function
\be \label{strip_example}
\Ztop=s_{\beta_1} s_{\beta_2} s_{\beta_3} s_{\beta_4} \frac{\{\beta_1 \beta_3 \}_{Q_1 Q_2} \{\beta_1 \beta_4 \}_{Q_1 Q_2 Q_3} \{\beta_2 \beta_3\}_{Q_2} \{\beta_2 \beta_4\}_{Q_2 Q_3}}{\{\beta_1 \beta_2^t\}_{Q_1} \{\beta_3^t \beta_4\}_{Q_3}} \,. 
\ee
Fixing the occurrence of transpositions requires a slightly more detailed analysis which is outlined in \cite{Iqbal:2004ne}.

The results for $\Ztop$ on the strip were used in \cite{Iqbal:2004ne} to study the behavior of the topological string partition function under flop transitions, and to prove the equality of Nekrasov's partition function and the corresponding topological string partition function for arbitrary rank $U(N)$ gauge group and matter content. In the next subsection, we wish to touch upon a more elaborate application, which was the subject of two papers \cite{Eynard:2010dh,Eynard:2010vd}. 

\newpage

\subsubsection{Application: $\Ztop$ via matrix models}
\begin{figure}[h]
\centering
\includegraphics[width=4cm]{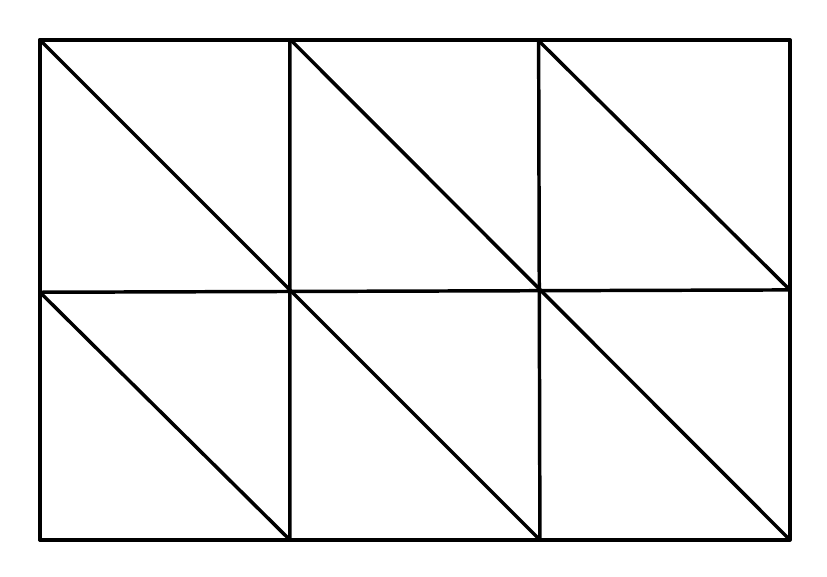}
\caption{A fiducially triangulated rectangle.}  \label{fig:triang_rect}
\end{figure}

The point of departure is the so-called BKMP conjecture, named after its authors \cite{Bouchard:2007ys}, which proposes an alternative method for computing the topological string amplitudes $F_g$ on a toric local Calabi-Yau geometry $X$, based on the topological recursion \cite{Eynard:2007kz}. The latter is an algorithm, inspired by matrix model computations, which assigns to any affine curve $\cC$ functions $\cF_g$ of the curve data. If $\cC$ coincides with the spectral curve of a matrix model, then the $\cF_g$ yield the coefficients of the logarithm of the partition function of the matrix model in a large $N$ expansion. These functions share many properties with topological string amplitudes $F_g$ \cite{Chekhov:2003gz,Eynard:2007hf}. This prompted the authors of \cite{Bouchard:2007ys} to conjecture that an adaptation of the topological recursion algorithm applied to the mirror curve of a local Calabi-Yau geometry $X$ should yield functions $\cF_g$ which {\it equal} the topological string amplitudes $F_g(X)$. To study this conjecture, Eynard and I set out to construct a matrix model whose partition function reproduces that of the topological string on an auspiciously chosen fiducial geometry $\xf$ \cite{Eynard:2010dh,Eynard:2010vd}. The proof of the BKMP conjecture for this geometry then reduces to demonstrating that the spectral curve of this matrix model coincides with the mirror curve of $\xf$. It is easy to show \cite{Eynard:2010vd} that any toric Calabi-Yau manifold can be related to a sufficiently large triangulated rectangle as depicted in figure \ref{fig:triang_rect} upon blowing up vertices and flopping a number of curves. As the effect of these two operations on the topological string amplitudes $F_g$ is well understood, proving the conjecture for such a fiducial geometry $\xf$ will thus quickly imply the general result.

In this subsection, we will sketch how to construct a matrix model that computes the topological string partition function on $\xf$. This is the content of \cite{Eynard:2010dh}. Computing the spectral curve of this matrix model, the subject of the follow-up work \cite{Eynard:2010vd}, requires elaborate matrix model technology which would lead us too far afield. To summarize the conclusion of \cite{Eynard:2010vd}, the potential of the matrix model we will propose lies outside of the validity of several theorems we require to compute the spectral curve. Upon proposing certain natural generalizations of these theorems, the mirror curve indeed emerges in a highly non-trivial fashion as the spectral curve of our matrix model. Elevating our arguments to a proof of the BKMP conjecture will require extending the necessary theorems, a task of independent interest. Since the publication of \cite{Eynard:2010vd}, a proof of the conjecture using different means has been proposed in \cite{Eynard:2012nj}.

We turn now to the derivation of the matrix model. We wish to convey a flavor of why a chain of matrices matrix model arises, and how matrix integrals can give rise to sums over partitions.

\begin{figure}[h]
\centering
\includegraphics[width=8cm]{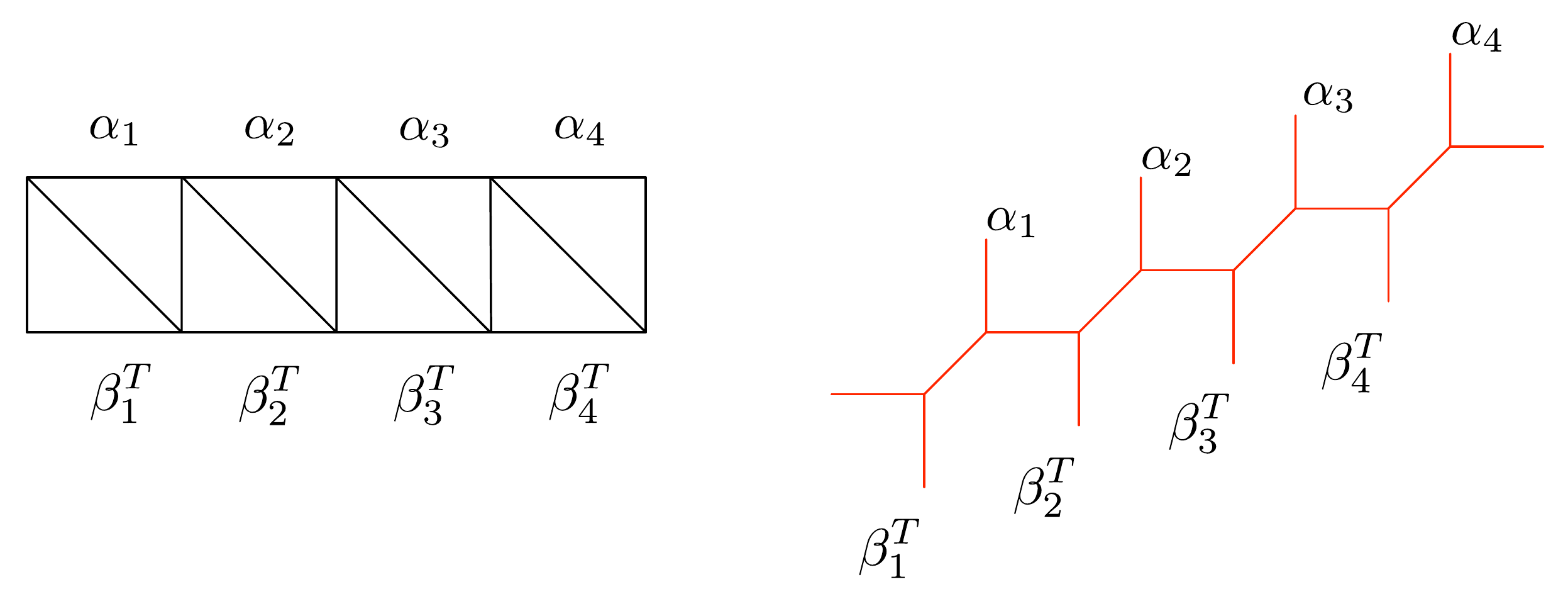}
\caption{A fiducial triangulation of the strip.}  \label{fig:fid_strip}
\end{figure}

\paragraph{The chain of matrices matrix model:} The topological string partition function $\zf$ on $\xf$ can be computed by gluing together the building blocks $\zs$ depicted in figure \ref{fig:fid_strip} by summing over the Young diagrams $\alpha_i$ and $\beta_i$.
The basic challenge is to rewrite these sums as matrix integrals. Following the rules developed in \cite{Iqbal:2004ne} and reviewed above, $\zs$ on a strip of $n+1$ boxes is given by
\begin{equation}
\zs (\alpha_0, \ldots, \alpha_n;\beta_0^T, \ldots, \beta_n^T)  \label{z_strip}
=   \prod_{i=0}^n \frac { s_{\alpha_i} s_{\beta^T_i}  }{ [\beta_i,\alpha_i^T]_{Q_{\beta_i,\alpha_i}}} \,\,\,\, {\prod_{i<j} [\alpha_i,\alpha_j^T]_{Q_{\alpha_i,\alpha_j}} \,\,\prod_{i<j} [\beta_i,\beta_j^T]_{Q_{\beta_i,\beta_j}} 
\over \prod_{i<j} [\alpha_i,\beta_j^T]_{Q_{\alpha_i,\beta_j}} [\beta_i,\alpha_j^T]_{Q_{\beta_i,\alpha_j}}
}    \,,
\end{equation}
where $Q_{\gamma, \delta}$ signifies the appropriate exponentiated K\"ahler parameter for the curve spanning between the vertices labeled by $\gamma$ and by $\delta$. As a first step towards the matrix model, we re-express $\zs$ in terms of the diagonal matrices
\be  \label{eigenvalues}
X(\gamma) = \diag(q^{h_1}, \ldots, q^{h_d}) \,, \quad h_k(\gamma) = \gamma_k - k + d + a_\gamma  \,,
\ee
with the parameters $a_\gamma$ encoding the K\"ahler parameters of the geometry \cite{Eynard:2010dh}. We have introduced an upper bound $d$ on the number of rows of the Young diagrams we consider. It can be taken to be arbitrarily large.\footnote{The spectral curve will depend only non-perturbatively on $d$ \cite{Eynard:2010vd}.}  In terms of the Vandermonde determinant
\be
\Delta(X) = \prod_{1\leq i<j\leq d} (X_j-X_i) \,,
\ee
and its generalization
\be
\Delta(X(\gamma),X(\delta)) = \prod_{i,j} (X_i(\delta)-X_j(\gamma)) \,,
\ee
$\zs$ can be written as
\ba
\lefteqn{\zs (\alpha_0,\dots,\alpha_n;\beta_0^T,\dots,\beta_n^T) =} \nn\\
&& \\
&=& {\prod_i \Delta(X(\alpha_i))\,\prod_{i<j}\Delta(X(\alpha_i),X(\alpha_j))\,\,\, \prod_i \Delta(X(\beta_i))\,\prod_{i<j}\Delta(X(\beta_i),X(\beta_j))
\over \prod_{i,j} \Delta(X(\alpha_i),X(\beta_j))} \nn\\
&& \times \exp \left[ -\frac{1}{g_s} \tr \, V\left(X(\alpha_i), X(\beta_i) \right) \right]  \,.
\ea
Here, $V$ is a complicated potential, the details of which can be found in \cite{Eynard:2010dh}.
By introducing larger matrices still \cite{Klemm:2008yu},
\be
X_1 =  \diag (X(\alpha_0), \ldots, X(\alpha_n))  \,, \quad X_2 =  \diag (X(\beta_0), \ldots, X(\beta_n)) \,, 
\ee
we can rewrite
\be
 {\prod_i \Delta(X(\alpha_i))\,\prod_{i<j}\Delta(X(\alpha_i),X(\alpha_j))\,\,\, \prod_i \Delta(X(\beta_i))\,\prod_{i<j}\Delta(X(\beta_i),X(\beta_j))
\over \prod_{i,j} \Delta(X(\alpha_i),X(\beta_j))} = {\Delta(X_1) \Delta(X_2)\over \Delta(X_1,X_2)}  \,. \label{vands}
\ee  
We next want to re-express (\ref{vands}) in terms of a certain set of normal matrices which we shall define below, rather than diagonal matrices. To this end, we introduce $N=(n+1)d$ additional integrals and write
\be  \label{enter_y}
 (-1)^{\binom{N}{2}} N!\, g_s^N\,\, {\Delta(X_1)\,\Delta(X_{2})\over \Delta(X_{1},X_2)} =
\int_{\mathbb R_+^N} dY    \mathop{{\det}}_{p,q}(e^{{-1\over g_s} (X_{1})_{p}\, (Y)_{q} })   \,\,\,   \mathop{{\det}}_{p,q}(e^{{1\over g_s} (X_2)_{p}\, (Y)_{q} }) \,,
\ee
with $(X)_p$ indicating the $p$-th diagonal entry of the matrix $X$. We have here used Cauchy's determinant formula
\be
\det \left( \frac{1}{x_i + y_j} \right)_{1 \le i < j  \le n} = \frac{\prod_{1 \le i < j  \le n} (x_j - x_i) (y_j - y_i)}{\prod_{i,j=1}^n (x_i + y_j)} \,.
\ee
Each determinant factor in (\ref{enter_y}) can be written as an Itzykson-Zuber integral \cite{ItzyksonZuber},
\be
 {\mathop{{\det}}_{p,q} (e^{x_p\,y_q})=  \Delta(X)\,\Delta(Y)}  \int dU\, e^{\tr\, X U Y U^\dagger} \,,
\ee
where $X$ and $Y$ are now arbitrary normal matrices with eigenvalues $x_p$, $y_q$ respectively. Substituting $U= U^\dagger_{X_1} U_{Y}$ and $U = U^\dagger_{X_2} U_{Y} $ respectively for the two determinants on the right-hand side of (\ref{enter_y}), and introducing
\be
M_i = U_{X_i} X_i U^\dagger_{X_i}  \,, \quad  R = U_{Y} Y U^\dagger_{Y} \,,
\ee
we arrive at a contribution
\be
\propto \Delta(M_1) \Delta(M_2) \Delta(R)^2 e^{\frac{1}{g_s}\tr\, (M_2 - M_1) R } 
\ee
per strip.

We thus see the general structure for the matrix model computing $\zf$ emerging: the Young diagrams above and below a strip, denoted by $\alpha_i$, $\beta_i$ in figure \ref{fig:strip}, are encoded in matrices $M_1$, $M_2$. Strips are glued by integrating over the corresponding $M_i$. Each strip also involves a matrix $R$, which only appears linearly in a contribution to the potential of the form $\tr\, (M_2 - M_1) R$. No other contributions to the potential involve products of different matrices. This is the general structure of a class of matrix models called chain of matrices.

\paragraph{From matrix integrals to sums over Young diagram:} Next, we need to constrain the eigenvalues of the matrices $M_i$ to be of the form (\ref{eigenvalues}). This is accomplished by choosing their integration domain as
\be
H_N(\Gamma_i) = \{ M=U\, \Lambda \, U^\dagger \, , \quad U\in U(N)\, , \,\,\, \Lambda={\rm diag}(\lambda_1,\dots,\lambda_N)\, \in \Gamma_i \} \,,
\ee
with $\Gamma_i$ the product of contours
\be
\Gamma_i = \prod_{j=0}^n\, (\gamma_{j,i})^d 
\ee
passing through the desired eigenvalues of the form (\ref{eigenvalues}). We then multiply the integrand of the matrix model with functions $f_i$, the explicit form of which can be found in \cite{Eynard:2010dh}, that exhibit simple poles at these eigenvalues, and constant residue. The integral over eigenvalues is thus replaced by a sum over terms of the form (\ref{eigenvalues}) for arbitrary positive integers $\gamma_k$.   Due to the symmetry of the integrand in the eigenvalues, we can, upon multiplying by the number of permutations, order them, such that the $\gamma_k$ once again reflect the number of boxes in the $k$-th row of a Young diagram $\gamma$. 

\newpage

\section{$\Ztop$ via holomorphic anomaly}  \label{section:hae}

The previous section unfolded mainly within the A-model version of the topological string, and built on target space considerations. We now turn to a computation method using B-model techniques which is firmly rooted in worldsheet considerations. We will hence be computing order by order in the string coupling.  This appears to be taking a step backwards compared to the results of the topological vertex, which are non-perturbative in the string coupling. What we gain, however, is the ability to compute all order results in the K\"ahler parameters. As a consequence, modularity properties will become manifest that are not visible at finite order.

\subsection{The BCOV holomorphic anomaly equations}   
 
The topological string amplitudes $F_g$ are naively, by BRST symmetry, holomorphic functions on the appropriate moduli space of a Calabi-Yau manifold $X$: they are defined as correlators of a two-dimensional twisted worldsheet theory integrated over the compactified moduli space of genus $g$ Riemann surfaces $\cM_g$. Derivatives with regard to anti-holomorphic variables lead to additional insertions in the correlator. As these are BRST exact, they give rise to contributions only from the boundary of the integration domain. In the case at hand, the boundary of $\cM_g$ describes degenerating Riemann surfaces. Pictorially, degenerations arise upon the pinching of cycles of the Riemann surface, giving rise to lower genus Riemann surfaces. A careful study of this phenomenon \cite{BCOV} gives rise to the holomorphic anomaly equations,
\ba   \label{hae} 
\bar{\partial}_{\bar{\imath }} F^{g}= \frac{1}{2}\bar{C}_{\bar{\imath }}^{jk}\big{(}D_jD_kF^{g-1} 
+\sum_{h=0}^{g-1}  D_jF^{h}D_kF^{g-h}\big{)} \,,  \quad g>1 \,.
\ea 
Here, $\bar C^{ij}_{{\bar \imath}}=e^{2{\cal K}}  G^{j\bar \jmath} G^{k\bar k} C_{\bar \imath\bar \jmath \bar k}$, where the K\"ahler potential ${\cal K}$ and the metric $G^{j\bar \jmath}$ on moduli space as well as the three-point function $C_{ijk}$ are special geometry data encoded in the genus 0 amplitude $F^0$ of the theory. The derivatives $D_i$ are covariant with regard to the appropriate bundles. The contributions on the right-hand side of this equation can easily be traced to the two possible ways in which a Riemann surface can degenerate in $\cM_g$: if the pinched cycle does not sever the Riemann surface in two, the resulting surface has genus reduced by one, giving rise to the first term on the right-hand side of (\ref{hae}). If on the other hand the surface becomes disconnected upon pinching, the resulting two components of genus $h$ and $g-h$ give rise to the second contribution to the anti-holomorphic derivative.

As the genera appearing on the right-hand side of the holomorphic anomaly equation are strictly smaller than the genus on the left-hand side, the equation gives rise to a recursion relating topological string amplitudes at different genera. The starting point of the recursion is the topological string amplitude $F^1$ at genus one, which satisfies its own holomorphic anomaly equation expressed purely in terms of special geometry data. Clearly, the recursion is not sufficient to fix the topological string amplitudes, as it contains no information about the purely holomorphic dependence of the genus $g$ amplitude. We will see below how this so-called holomorphic ambiguity can be addressed by providing appropriate boundary conditions.

\subsection{The generalized holomorphic anomaly equations}  
We will here be interested in the refined topological string briefly introduced in subsection \ref{s:beyond_gae}, in particular with an eye towards the applications in section \ref{s:agt}. As a worldsheet definition of the $\Omega$-deformation is still lacking, a derivation of recursion relations for the refined amplitudes following the reasoning of \cite{BCOV} presented above is currently not available. In its place, a simple generalization of the equations (\ref{hae}) was conjectured in \cite{HK3}. These equations have by now passed numerous checks \cite{HK3, KW1,Huang:2011qx}. They are given by 
\begin{eqnarray} \label{gen_hol_ano} 
\bar{\partial}_{\bar{i}} F^{(n,g)}= \frac{1}{2}\bar{C}_{\bar{i}}^{jk}\big{(}D_jD_kF^{(n,g-1)} 
+{\sum_{m,h} }^{\prime}  D_jF^{(m,h)}D_kF^{(n-m,g-h)}\big{)} \,, \quad n+g>1  \,.
\end{eqnarray} 
The prime on the sum indicates omission of the summands $(m,h)=(0,0)$ and $(m,h)=(n,g)$. The first term on the right-hand side is absent at $g=0$.  

Note that these equations are reminiscent of the holomorphic anomaly equations for the topological string with insertions \cite{BCOV}, suggesting that the $\Omega$-deformation may correspond to an appropriate operator insertion in the correlator defining the generalized amplitudes. In \cite{Huang:2011qx}, we briefly discuss the dilaton as a possible candidate for such an insertion. This proposal deserves further study.

\subsection{The holomorphic anomaly equations in the rigid limit} 
\label{directintegration} 
We will here be interested in applying the holomorphic anomaly equations to rigid $\cN=2$ theories which arise upon $\Omega$-deformation of string theory on non-compact Calabi-Yau manifolds \cite{Huang:2011qx}. We will furthermore specialize to gauge theories, though the formalism applies equally well to arbitrary local geometries and without taking a field theory limit.  We will see that with minimal assumptions regarding the modularity of the result, we can integrate the holomorphic anomaly equations exactly to any desired order in $g_s$ and $s$. 

Before we specialize to the special geometry of a genus one curve (this is the target space in question, not to be confused with genus one worldsheets), most of the discussion in this subsection can easily be presented in greater generality, simply by increasing the number of variables and indices. The main technical simplification that occurs when restricting to genus one is that the solutions of the Picard-Fuchs equations yielding the relevant periods required to specify the special geometry can be written down explicitly.

\newpage

\subsubsection{The special geometry of Seiberg-Witten theory}   \label{s:sw_rev}
Seiberg-Witten theories \cite{SW1,SW2} are four-dimensional gauge theories exhibiting $\cN=2$ supersymmetry. Here, we will focus on the gauge group $SU(2)$. The fields are organized in vector and hypermultiplets. The $\cN=2$ vector multiplet consists of an $\cN=1$ chiral and an $\cN=1$ vector multiplet. Much of the physics is determined by the complex scalar $\phi$ in the chiral multiplet. The potential for $\phi$ exhibits a flat direction. The corresponding vacuum expectation value $u= \frac{1}{2} \Tr(\phi^2)$ hence parametrizes a moduli space, which coincides with the moduli space of the topological string (in the field theory limit). This vacuum expectation value breaks the gauge symmetry from $SU(2)$ to $U(1)$. The low-energy two derivative effective action of the theory (before coupling to a gravitational background) is entirely encoded in terms of a holomorphic quantity in $\phi$ called the prepotential $F^{(0,0)}$. Much of the power of the Seiberg-Witten approach to solving these theories stems from identifying contributions to the action with geometric quantities on moduli space. Due to the presence of $\cN=2$ supersymmetry, the moduli space is governed by {\it special geometry}. We will describe the implications of this structure in the following.

Let us first determine the geometric data that enter into the holomorphic anomaly equations (\ref{hae}). In terms of distinguished flat coordinates $t$ which can always be defined locally, the metric on moduli space is given by
\be \label{PW_metric} 
G_{t \bar{t}} = 2 \del_t \del_{\bar{t}} \re (\bar{t} \del_t F^{(0,0)}) \,.
\ee 
This metric is visibly K\"ahler, with the corresponding K\"ahler potential $\cK=\re (\bar{t} \del_t F^{(0,0)})$. The three-point functions $C_{ijk}$ occurring in (\ref{hae}) are determined by triple derivatives of the prepotential. In the one-dimensional case that we are considering here, 
\be 
C_{ttt} = \frac{\del^3 F^{(0,0)}}{\del t^3}  \,.
\label{general3pointcoupling} 
\ee 
The geometry on moduli space is hence entirely determined by the prepotential $F^{(0,0)}$ and the appropriate choice of flat coordinates. This data can be encoded \cite{SW1,SW2} in terms of a family of Riemann surfaces $\cC_1(u)$ of genus one (higher rank groups require curves of higher genera) parametrized by the modulus $u$ of the theory, as well as a meromorphic (1,0)-form $\lambda$, the so-called Seiberg-Witten differential. This one-form is constrained to satisfy $\frac{d \lambda}{du} = \omega$, with $\omega$ the unique (up to scaling) holomorphic one-form on $\cC_1$. The periods of $\lambda$ along one-cycles $(\Sigma_A, \Sigma_B)$ furnishing a symplectic basis of $H_1(\cC_1,\IZ)$ determine the flat coordinate $a$ and its dual $a_D$,\footnote{In massive theories, the Seiberg-Witten differential exhibits residues proportional to the masses, and the integration contour must be specified.}
\begin{equation}  \label{per_lambda}
a=\oint_{\Sigma_A} \lambda, \qquad a_D=\oint_{\Sigma_B} \lambda \,. 
\end{equation} 
This definition fixes $a$ uniquely as a flat coordinate in the vicinity of $u \rightarrow \infty$, the weak coupling point of the gauge theory, upon fixing boundary conditions
\be  \label{bc_for_a}
a \sim c_0 \sqrt{\frac{u}{2}}   \quad \mbox{at} \quad u\rightarrow \infty\,.
\ee
The constant $c_0$ is equal to one or two, depending on whether the gauge theory contains fundamental matter or not. Away from this point, the periods are multi-valued functions of $u$. We will refer to their values on any given branch as $\ttd$. Locally, the function $t(u)$ can be inverted, and $t_D$ thereupon expressed as a function of $t$. 

The physical interpretation of flat coordinates is most evident by considering the expression for the central charge and the complex gauge coupling of the theory. The central charge associated to a BPS particle of electric, magnetic, and $U(1)$ flavor charge $(n_e, n_m, S_i)$ is given by
\be
\label{centralcharge} 
Z= n_e a + n_m a_D + \sum_i S_i \frac{m_i}{\sqrt{2}} \,.
\ee 
By $\cN=2$ supersymmetry, the mass of a BPS particle carrying these charges is determined by the central charge via $m = |Z|$. 

A hallmark of $\cN=2$ theories is that the gauge coupling and theta angle can be combined into a complex gauge parameter $\tau$,
\be \label{tau} 
\tau =  \frac{1}{c_0} \left( \frac{\theta}{\pi} + \frac{8 \pi i}{g^2} \right) \,.
\ee
The effective infrared gauge coupling of the effective low energy theory is determined in terms of $\ttd$ via
\be
\tau = \frac{dt_D}{dt} \,.
\ee
Note that via the relation of $\lambda$ to the holomorphic one-form on $\cC_1$,
\be
\tau = \frac{d t_D}{du} / \frac{d t}{du} = \dfrac{\int_{\Sigma_B} \omega}{\int_{\Sigma_A} \omega} \,. 
\ee
As the ratio of two symplectically dual periods of $\omega$ takes values in the upper half-plane, the positivity of the effective gauge coupling is thus manifest in this formalism.

The structure (\ref{per_lambda}) permits the computation of the prepotential due to the special geometry relation
\be \label{dual_period}
t_D = -\frac{c_0}{2 \pi i}  \frac{\partial F^{(0,0)}}{\partial t}  \,.
%\quad \mbox{with} \quad
%\begin{cases} 
%c_0=1 \quad \mbox{theory with fundamental matter} \,,\\ 
%c_0=2 \quad \mbox{theory without fundamental matter}  \,.
%\end{cases} 
\ee
Computing the prepotential hence requires determining the periods of the meromorphic differential $\lambda$, expressing $t_D$ as a function of $t$, and integrating once, or determining $\tau$ as a function of $t$, and integrating twice.

\subsubsection{Computing periods of $\lambda$}  \label{comp_periods}
The general path towards computing the periods of the Seiberg-Witten differential $\lambda$ (as well as the corresponding problem of finding the periods of the holomorphic (3,0) form $\Omega$) proceeds via solving the Picard-Fuchs differential equations that these periods satisfy. In the case at hand, for which the relevant curve is elliptic, a general formula is available to express the relevant period of the holomorphic one-form $\omega$. The corresponding period of $\lambda$ can then be obtained via integration. Before introducing this formula, we will rapidly review some basic facts about elliptic curves.

Upon suitable variable definition, the elliptic curve $\cC_1$ can generically be put in Weierstrass form
\begin{eqnarray}  \label{curve_equation} 
y^2 = 4x^3 -g_2(u) x- g_3(u) \,.
\end{eqnarray}  
The complex structure of this curve is specified by the ratio of the periods of the holomorphic one-form. Not coincidently, this ratio is denoted as $\tau$. The symplectic basis $(\Sigma_A, \Sigma_B)$ with regard to which the periods are computed is only defined up to an action of the group $SL(2,\IZ)$, and two $\tau$ parameters related by the induced $SL(2,\IZ)$ transformation determine the same complex structure. Due to this degeneracy, a more convenient parametrization of the complex structure is given by the $J$-invariant of the curve, 
\begin{eqnarray}  
\label{jdefu}  
J= \frac{g_2(u)^3}{\Delta(u)}\,.
\end{eqnarray} 
$\Delta$ here denotes the discriminant of the curve, 
\begin{eqnarray} 
\Delta(u) = g_2(u)^3-27g_3(u)^2 \,. 
\end{eqnarray} 
The relation between $J$ and $\tau$ is established by the formula
\begin{eqnarray}  
\label{jdeftau} \label{relation1} 
J(\tau)= \frac{E_4(\tau)^3}{E_4(\tau)^3-E_6(\tau)^2} \,,
\end{eqnarray} 
where $E_4$ and $E_6$ are Eisenstein series, modular forms of $SL(2,\IZ)$ of weight 4 and 6 respectively. The modular invariance of $J$ is manifest in this formula.

The curve is singular over points on moduli space at which the $J$-invariant is infinite. Aside from the weak coupling point at $u \rightarrow \infty$, this occurs at zeros of the discriminant. Physically, singularities in the interior of moduli space correspond to points at which particles are becoming massless \cite{SW1,SW2}. Generically, periods undergo monodromy upon circling singularities. According to the theory of Picard-Fuchs equations applied to elliptic curves, a unique period (corresponding to a choice of cycle in $H_1(\cC_1, \IZ)$) exists at each such singularity that does not undergo monodromy, the so-called constant period. This is the period we call $t$ in general, and $a$ in the particular case of the singular point at $u \rightarrow \infty$. Up to normalization, this period is uniquely determined to be
\begin{eqnarray} \label{nonlogperiod} \label{relation2} 
\frac{dt}{du} =c_1 \sqrt{\frac{g_2(u)}{g_3(u)} \frac{E_6(\tau)}{E_4(\tau)}}=3^\frac{1}{4} c_1 \sqrt[4]{\frac{E_4(\tau)}{g_2(u)}}  \,.
\end{eqnarray}
To obtain the period $a$, the physical boundary condition (\ref{bc_for_a}) must be imposed.

Given these formulae, we can hence compute the prepotential $F^{(0,0)}$ following the second method outlined at the end of subsection \ref{s:sw_rev}: we first use (\ref{jdefu}) and (\ref{jdeftau}) to express $\tau$ (i.e. an appropriate representative of the $SL(2, \IZ)$ orbit of the complex structure parameter) in terms of the variables parametrizing the Seiberg-Witten curve of the theory. Generically, this will yield $\tau$ in a power series expansion in these variables. Substituting $\tau(u)$ into (\ref{nonlogperiod}) allows us to determine $t(u)$. Inverting this relation and substituting yields $\tau(t)$, from which we can finally compute the prepotential by integrating twice,
\be  \label{prepot_int_twice}
F^{(0,0)}  \sim \int^t dt \int^t dt \, \tau \,.
\ee

\subsubsection{Modularity}  \label{s:modularity}
The direct integration of the holomorphic anomaly equations \cite{GKMW} relies on the following two observations. Firstly, the equations are formulated with regard to local variables on moduli space. The effective coupling parameter $\tau$ is such a variable. We have seen that as a function of the global coordinate $u$, $\tau$ is multivalued; it transforms under the monodromy group $\Gamma \subset SL(2,\IZ)$ upon circling singularities in the moduli space. As physical quantities such as the refined topological amplitudes $F^{(n,g)}$ for $n+g>1$ (for the three remaining cases, only the derivatives have physical significance) should be unique functions of $u$ (otherwise the true moduli space of the theory would be a cover of the $u$-plane), we can conclude that such quantities expressed as functions of $\tau$ must be invariant under the monodromy group $\Gamma$. Secondly, the point of departure for formulating the holomorphic anomaly equations is the non-holomorphicity of the $F^{(n,g)}$. Using the propagator method introduced in \cite{BCOV}, one can argue that the non-holomorphicity of these quantities can be captured by expressing them as polynomials in $\frac{1}{\tau_2}$, $\tau_2 = \im \,\tau$, with holomorphic functions as coefficients. Combining these two observations, it is natural to attempt to express the $F^{(n,g)}$ in terms of almost holomorphic modular forms \cite{Zagier} for the monodromy group $\Gamma$. These are modular forms under $\Gamma \subset SL(2,\IZ)$  that are polynomials in $\frac{1}{\tau_2}$ with coefficients that are holomorphic functions of $\tau$. They form the ring $\widehat{M}(\Gamma)$. The simplest representative of this class of functions is the modular completion $\hat{E}_2$ of the second Eisenstein series $E_2$,
\be \label{Ehat}
\hat{E}_2(\tau, \bar{\tau}) = E_2(\tau) - \frac{3}{\pi \tau_2} \,. 
\ee 
In fact, this function can be used to provide a second definition of almost holomorphic modular forms: they are polynomials in $\hat{E}_2$ with coefficients that are holomorphic modular forms (the equivalence of these two definitions is the content of Prop. 20 in \cite{Zagier}). A closely related class of functions is obtained by mapping $\hat{E}_2$ to $E_2$, giving rise to the ring $\widetilde{M}(\Gamma)$ of quasi-modular forms. They exhibit holomorphic dependence on the modular parameter, yet transform in a complicated fashion due to the modular anomaly of $E_2$. As the rings $\widehat{M}(\Gamma)$ and $\widetilde{M}(\Gamma)$ are isomorphic, we will use the two notions interchangeably in the following without further comment. 

Taking the anti-holomorphic derivative in (\ref{gen_hol_ano}) with regard to the coordinate $\bar{\tau}$ and invoking (\ref{Ehat}), and taking the derivatives on the right-hand side of the equation with regard to the flat coordinate $a$, yields the equation \cite{HK2} 
\begin{eqnarray}  \label{hae_hol} 
24 \frac{\del F^{(n,g)} }{\del {\hat{E}_2}} =   c_0 \big{(} \frac{\del^2 F^{(n,g-1)}}{\del a^2} 
+{\sum_{m,h} }^{\prime} \frac{\del F^{(m,h)}}{\del a} \frac{\del F^{(n-m,g-h)}}{\del a}    \big{)}. 
\end{eqnarray} 
To obtain expressions valid globally on moduli space, we replace the $a$ with $u$ derivatives and arrive at
\ba
24 \frac{\partial F^{(n,g)} }{\partial X} &=& c_0 \frac{g_2(u)}{g_3(u)} \frac{E_6}{E_4} \Big[ \left(\frac{d u}{d a}\right)^2 \frac{\del^2 F^{(n,g-1)}}{\del u^2} +\frac{d^2 u}{d a^2} \frac{\del F^{(n,g-1)}}{\del u} \nn \\ 
&& +\left(\frac{d u}{d a}\right)^2 {\sum_{m,h} }^{\prime} \frac{\del F^{(m,h)}}{\del u} \frac{\del F^{(n-m,g-h)}}{\del u} \Big] \,.   \label{hol_an_X}  
\ea 
We have here introduced the variable
\begin{eqnarray} 
X=\frac{g_3(u)}{g_2(u)} \frac{{E}_2(\tau)E_4(\tau)}{E_6(\tau)} \,.
\label{Xdef} 
\end{eqnarray} 
This proves computationally convenient, as the derivatives occurring in (\ref{hol_an_X}), with recourse to (\ref{nonlogperiod}) and upon invoking the Ramanujan identities, have handy expressions in terms of low order polynomials in $X$.
%\be  \label{d2uda2} 
%\frac{d^2 u}{d a^2}  = \frac{1}{\Delta} \frac{g_3(u)}{g_2(u)} \frac{E_4}{E_6} \,p_1(X)  \,,
%\ee 
%with $p_1(X)$ a first order polynomial in $X$ with coefficients that are polynomials of derivatives of $g_2(u)$ and $g_3(u)$. 
Taking the $n+g=1$ amplitudes, 
\begin{eqnarray}  
F^{(0,1)} &=& -\frac{1}{2} \log (G_{u\bar u} |\Delta|^\frac{1}{3})   \label{genus1a}\,, \\ 
F^{(1,0)} &=& \frac{1}{24}\log (\Delta) \label{genus1b}\,,  
\end{eqnarray} 
as point of departure, it is easy to derive the general form of the amplitudes $F^{(n,g)}$ via induction to be
\begin{equation} \label{generalformfng} 
F^{(n,g)}=\frac{1}{\Delta^{2(g+n)-2}(u)} \sum_{k=0}^{3g+2n-3} X^k p^{(n,g)}_k(u) \,.
\end{equation} 
Here, $p^{(n,g)}_k(u)$ are polynomials in derivatives of $g_2(u)$ and $g_3(u)$. The holomorphic anomaly equations fix all of these, with the exception of $p_0^{(n,g)}$, the holomorphic ambiguity alluded to above.  Requiring the finiteness of $F^{(n,g)}$ as $u \rightarrow \infty$ constrains the degree of $p^{(n,g)}_0(u)$ as polynomials in $u$, resulting in a finite number of coefficients that need to be fixed at each $(n,g)$.

% From this discussion, one might be tempted to conclude that the symmetry group of the theory is the full modular group $SL(2,\IZ)$. This is incorrect. The holomorphic anomaly equations as written in (\ref{hol_an_X}) depend on both UV parameters ($u$, potentially bare masses, and the UV coupling $\tau_{uv}$ or the dimensional transmutation scale $\Lambda$) and IR parameters (the effective coupling $\tau$, the argument of the Eisenstein series contained in $X$). To draw conclusions regarding the symmetry group, one should re-express it fully in terms of IR parameters. That this reduces the symmetry group to a subgroup of $SL(2,\IZ)$ can be seen explicitly e.g. in the massless asymptotically free cases of $SU(2)$ gauge theory with $N_f<4$ flavors, in which $u$ can be obtained explicitly as a function of $\tau$ (see e.g. \cite{HK2}) and proves to be modular only under the monodromy group $\Gamma \subset SL(2,\IZ)$. 

% The leading power of $X$ can be lower in theories where the leading $p^{(n,g)}_k$ vanish identically -- this happens in the conformal cases studied in this paper, massless $N_f=4$ and $N=4$. The above derivation fails as $\frac{d\tau}{da}=0$ in the conformal cases. Likewise, the leading negative power of the discriminant can be lower if all $p^{(n,g)}_k$ contain powers of the discriminant, as turns out to be the case for mass deformed $N=4$. 

\subsubsection{BPS states and fixing the holomorphic ambiguity} 
\label{fixingtheambiguity}  
The holomorphic anomaly equations manifestly do not contain sufficient information to fix the purely holomorphic part of the amplitudes $F^{(n,g)}$. We fix these with recourse to the interpretation of $\Ztop$ as a counting function of BPS particles, on which the parametrization (\ref{ztop_gv}) and its refinement relies \cite{Haghighat:2008gw,Huang:2010kf,Huang:2011qx}. Note that unlike the holomorphic anomaly equations, this input follows from target space, not worldsheet considerations. 

The argument relies on the fact that, for generic masses of the matter fields in the theory, a single BPS particle is becoming massless at a singularity of the curve $\cC_1$ at which its discriminant vanishes. The constant period (\ref{nonlogperiod}) vanishes at such points. Expanding the contribution to the BPS counting function stemming from a single particle around $t=0$ yields the result
\be  \label{gap} 
F^{(n,g)} = \frac{N^{(n,g)}}{t^{2(g+n)-2}} + \mathcal{O} (t^0) \,,
\ee 
where the $N^{(n,g)}$'s are fixed fractions. We emphasize that this result does not require further knowledge of the BPS spectrum (this would be akin to knowing $\Ztop$), as each singularity, for a generic choice of mass parameters, is due to a single particle. The absence of subleading poles in $t$ is referred to as the gap condition ~\cite{HK1,Huang:2006hq}.  A counting argument demonstrates that the constraints imposed by this condition at all singularities are sufficient to fix the holomorphic ambiguity ~\cite{Haghighat:2008gw,Huang:2011qx}.

\subsection{The $SU(2)$ $\cN=4$ theory and its deformation to $\cN=2^*$}  \label{s:n2_holanomaly}
Coupling $\cN=2$ $SU(2)$ gauge theory to a massless adjoint matter multiplet enhances the supersymmetry of the theory to $\cN=4$. The resulting gauge theory is superconformal. This modifies the nature of the parametrization of the Seiberg-Witten curve, compared to asymptotically free cases. For the latter, the ultraviolet gauge coupling can be replaced, via dimensional transmutation, by a scale $\Lambda$. Demanding that the Seiberg-Witten curve remain finite both in the massless and the $\Lambda \rightarrow 0$ limit constrains its dependence on $\Lambda$ to be polynomial, for dimensional reasons. By contrast, an arbitrary holomorphic dependence of the curve on the ultraviolet gauge coupling $\tau_{UV}$ is a priori allowed in the case of superconformal theories. The theory even upon mass deformation (now referred to as $\cN=2^*$) exhibits S-duality with regard to this parameter, as we will demonstrate. A further particularity of the superconformal case is that the maximal none-vanishing power of $X$ in $F^{(n,g)}$ is $g+n-1$ rather than $2g+2n-3$ as in (\ref{generalformfng}). In this sense, the holomorphic anomaly is weaker for such theories.

\subsubsection{The Seiberg-Witten curve and the UV gauge coupling}   \label{s:n2_star_curve}
The Seiberg-Witten curve of $\cN=2^*$ was obtained in \cite{SW1}. It is given by
\begin{equation}  
 y^2=(x-e_1 u -\frac{1}{4} e_1^2 \, m^2)  (x-e_2 u -\frac{1}{4} e_2^2 \,m^2)(x- e_3 u -\frac{1}{4} e_3^2 \,m^2) \,.  
\label{N=4curve1} 
\end{equation} 
The $e_i(\tau)$ are the so-called half-periods of the Weierstrass $\wp$-function. Their occurrence in the Seiberg-Witten curve is natural when one considers the massless limit of the theory. The enhanced $\cN=4$ supersymmetry in this limit excludes quantum corrections to the prepotential. The ultraviolet and infrared gauge coupling hence coincide. Consequently, the complex structure of the curve (which, as we reviewed above, is identified with the $SL(2,\IZ)$ class of the infrared coupling) cannot depend on $u$. The Seiberg-Witten curve for this theory can therefore be identified with the generic elliptic curve of complex structure $\tau$ obtained from the relation
\be
\wp'^2 = 4 \wp^3 - g_2 \wp - g_3 
\ee
by setting $y = \wp'$, $x= \wp$. The roots in $x$ of the right-hand side coincide with $\wp$ evaluated at the zeros $\omega_i$ of $\wp'$, the two-torsion points of the torus. This is the definition of the half-periods, $e_i = \wp(\omega_i)$.

At finite mass, the prepotential does receive instanton corrections. The argument of the half-periods in (\ref{N=4curve1}) is now identified as the ultraviolet gauge coupling $\tau_{uv}$ of the theory, and no longer coincides with the complex structure of the curve, which exhibits $u$ and $m$ dependence.

\subsubsection{Calculating the amplitudes from the curve} \label{N=4amplitudesfromcurve} 
As we mentioned above, the $\cN=4$ supersymmetry that the $\cN=2^*$ theory exhibits in the massless limit rules out instanton corrections, such that that the prepotential is uncorrected and given by  
\begin{equation}  
F^{(0,0)} \sim \int^a da \int^a da \,\tau=\frac{1}{2} a^2 \,\tau \,. 
\label{fclassic} 
\end{equation} 
Following the steps outlined in section \ref{comp_periods}, the infrared coupling in the massive theory can be computed to be  
\be  \label{f00forN=4}  
2 \pi i\,  \tau = \log q  + 2 \log \frac{m^2 + 2 a^2}{2a^2}  + 6 \frac{m^4}{a^4}\, q + \frac{3 m^4 (24 a^4+80 a^2 m^2 +35m^4)}{4 a^8}\, q^2+ O(q^3)  \,,
\ee 
with $q= e^{2\pi i \tau_{uv}}$. The prepotential (\ref{prepot_int_twice}) is again obtained by integrating twice with regard to $a$. 

The amplitudes at $n+g>1$ are obtained by following the strategy outlined above, with a slight modification required \cite{Huang:2011qx} as the discriminant of the curve (\ref{N=4curve1}) is a perfect square: as a consequence, the discriminant vanishes to order two at all singularities of the curve, even though a single particle is becoming massless at these points. It is natural to identify this property as the physical manifestation of the presence of adjoint, rather than fundamental, matter in this theory. We obtain
 \begin{equation} \label{expansionY} 
F^{(n,g)}=\frac{1}{\tilde{\Delta}^{2(g+n)-2}(u)} \sum_{k=0}^{3g+2n-3} Y^k p^{(n,g)}_k(u) \,, 
\end{equation} 
where  
\be 
Y = (e_2-e_1) X 
\ee 
and $\tilde \Delta$ is essentially the square root of the discriminant,  
\be
\tilde \Delta = (4 u -e_1 m^2) (4u-e_2 m^2) (4u-e_3 m^2) \,. 
\ee 
To convey an impression of the form of these results, we here quote the amplitudes for $n+g=2$:
\ba
p^{(2,0)}_0&=&\frac{37  {E_4}^3 m^{10}-11232  {E_4}^2 m^6 u^2-96  {E_4} \left(7  {E_6} m^8 u+2376 m^2  u^4\right)-4  {E_6} m^4 \left(13  {E_6} m^6+20736 u^3\right)}{116640} \,, \nn\\ \nn 
p^{(2,0)}_1&=&  -\frac{\left( {E_4} m^4-144 u^2\right)^2}{432}  \,, \\ \nonumber  
p^{(1,1)}_0 &=& \frac{m^2 \left(- {E_4}^3 m^8+216  {E_4}^2 m^4 u^2+6  {E_4} \left( {E_6} m^6 u+1728 u^4\right)+ {E_6} m^2 \left( {E_6} m^6+2592 u^3\right)\right)}{2430} \,,\\ \nn 
p^{(1,1)}_1 &=&  \frac{1}{108} \left(5  {E_4}^2 m^8+288  {E_4} m^4 u^2+96  {E_6} m^6 u+20736 u^4\right) \,, \\ \nn 
p^{(1,1)}_2 &=& \frac{1}{2} \left(144 m^2 u^2- {E_4} m^6\right)  \,, \nn
\ea
\ba
p^{(0,2)}_0&=&\frac{m^2 \left(4  {E_4}^3 m^8+216  {E_4}^2 m^4 u^2+18  {E_4} \left(7  {E_6} m^6 u-5184 u^4\right)+ {E_6} m^2 \left( {E_6} m^6-14688 u^3\right)\right)}{43740}  \,, \nn \\ \nn 
p^{(0,2)}_1&=&  - \frac{1}{54} \left( {E_4}^2 m^8+144  {E_4} m^4 u^2+24  {E_6} m^6 u\right) \,, \\ \nn 
p^{(0,2)}_2&=&\frac{3}{2} m^2 \left( {E_4} m^4-144 u^2\right)  \,, \nn\\   
p^{(0,2)}_3 &=&  -45 m^4  \,. 
\ea

Note that the half-periods $e_i$ assemble to the Eisenstein series $E_4$ and $E_6$ in these polynomials. They are functions of the ultraviolet coupling $\tau_{uv}$. In contrast, the Eisenstein series entering in the definition of the variable $X$ are functions of the tau parameter of the Seiberg-Witten curve, the infrared coupling of the gauge theory.
 
\subsubsection{S-duality} \label{s:s-duality}
The results we have obtained are invariant under the monodromy group $\Gamma$, which acts on the infrared coupling $\tau$. In addition, the $\cN=2^*$ theory has an S-duality symmetry under the action of $SL(2,\IZ)$ on $\tau_{uv}$, which induces the same action on $\tau$. Expressing $u$ as a function of $\tau$ and $\tau_{uv}$ \cite{Huang:2011qx},
\be  \label{u_n4} 
u = \frac{m^2}{4} \frac{\aaa^2 (\BBB-\CCC) + \bbb^2(\CCC-\AAA)+\ccc^2(\AAA-\BBB)}{ \aaa (\BBB-\CCC) + \bbb(\CCC-\AAA)+\ccc(\AAA-\BBB)} \,,
\ee 
we read of the modular weight of $u$ under S-duality to be 2. The half-periods $e_i$ transform as weight 2 modular forms under the subgroup $\Gamma(2)$ of $SL(2,\IZ)$ consisting of elements of $SL(2,\IZ)$ which are equal to the identity element modulo 2.\footnote{More on this subgroup in section \ref{s:four_pt}.} Under the action of the full modular group, they are in addition permuted amongst themselves. 

Assembling all of these transformation properties, we can check the explicit S-duality invariance of our results.

\subsubsection{Expressing the $\cN=2^*$ theory in terms of infrared variables} \label{s:hae_ir}

The exact results we obtain for the refined amplitudes $F^{(n,g)}$ for $n+g>1$ are expressed as functions of a redundant set of variables $(u,\tau_{uv}, \tau)$. We can invert (\ref{f00forN=4}) to obtain $\tau_{uv}(\tau, a)$ and then invoke (\ref{u_n4}) to express $F^{(n,g)}$ purely in terms of the infrared variables $\tau$ and $a$. This is the form in which we will reproduce the amplitudes in section \ref{s:torus_one_pt}, as a power series in $\frac{m}{a}$. In the massless limit, this procedure can be performed exactly (as $\tau = \tau_{IR}$) and yields e.g.
\begin{equation} \label{results_N4_hae}
 F^{(2,0)}=\frac{E_2}{3 \cdot 2^6 a^2} \,,\qquad  F^{(1,1)}=-\frac{E_2}{3 \cdot 2^4 a^2} \,,\qquad F^{(0,2)}=0 \,,   
\end{equation} 
\ba 
&&F^{(3,0)}=-\frac{1}{2^{9} 3^2 5 a^4}\left(5E_2^2+13E_4\right) \,,   \quad F^{(2,1)}=\frac{1}{2^{8} 3^2 5 a^4}\left(25 E_2^2+29E_4\right)\,, \nonumber \\
&&F^{(1,2)}=-\frac{1}{2^{6} 3 \cdot 5 a^4}\left(5 E_2^2+E_4\right) \,,  \quad  F^{(0,3)}=0\,. \nn
\ea

\subsection{The $SU(2)$ $N_f=4$ theory}
The discussion of $SU(2)$ gauge theory with the $N_f=4$ fundamental flavors is in many regards similar to that of $\cN=2^*$. The full details can be found in \cite{Huang:2011qx}. Here, we only wish to highlight some differences. While both theories are superconformal and exhibit the same Seiberg-Witten curve in the massless limit, they are nevertheless not identical even at this level. This has two causes. The first is physical: the degenerate roots of the discriminant in the $N_f=4$ theory are indeed due to multiple particles becoming massless at these singularities. Upon generic mass deformation, all zeros become first order. The second is related to the choice of the ultraviolet coupling. As emphasized in \cite{Argyres:2007cn}, this reflects a choice of coordinates on the moduli space of marginal couplings of the theory, and is not canonical. To match the instanton results of Nekrasov, a different choice of ultraviolet coupling must be made in the $N_f=4$ and the $\cN=2^*$ theory: in the case of $\cN=2^*$, the argument of the half-periods $e_i$ appearing in the Seiberg-Witten curve is identified with $\tau_{uv}$; the appropriate choice in the $N_f=4$ theory is \cite{GKMW}
\be
e^{2 \pi i \tau_{uv}}=\frac{e_3 - e_2}{e_1 - e_2} \,.
\ee
While there is a systematics underlying this choice which we shall touch upon in section \ref{s:four_pt}, a first principles justification is still lacking.

\newpage

\section{AGT and $\Ztop$}  \label{s:agt}

In \cite{AGT}, a remarkable conjecture was put forward relating four dimensional superconformal $\cN=2$ $SU(2)$ quiver gauge theories to two dimensional Liouville conformal field theory. This conjecture goes by the name of the AGT correspondence, the initials of the authors of \cite{AGT}. The underlying intuition stems from considering an M5 brane on a six dimensional manifold which is the cross product of the four dimensional spacetime of the gauge theory and the two dimensional Riemann surface on which the conformal field theory is defined. Four dimensional gauge theories that can be obtained in this fashion are referred to as theories of class $\cS$ \cite{Gaiotto:2009hg}. By taking the volume of one or the other factor to be small, distinguished quantities in the effective M5 brane theory can be calculated in either a four dimensional or a two dimensional effective theory, and the corresponding observables of the four dimensional and two dimensional theory must hence coincide. As the M5 brane theory is only poorly understood, this reasoning furnishes useful intuition (also for analogous conjectures based on other partitions of six), but does not provide a detailed dictionary, or qualify as a proof of the correspondence. In \cite{Vartanov:2013ima}, a proof has been proposed based on the argument that the respective objects in two and in four dimensions are solutions to the same Riemann-Hilbert problem.

The bridge between the AGT correspondence and the topological string in the field theory limit is established via geometric engineering, which we already encountered in section \ref{s:geom_eng}, as the instanton partition functions of the superconformal field theories in question can be computed via the field theory limit of the topological string on appropriate geometries. Studying the topological string from the vantage point of this correspondence promises to be fruitful for at least two reasons. For one, $\Ztop$ on different geometries maps to different observables in the same conformal field theory, a unifying perspective. And secondly, whether via the original definition (\ref{ztop_basic}) or the index definition leading to (\ref{ztop_gv}), $\Ztop$ is defined as a formal power series, either in the string coupling or K\"ahler parameters. In contrast, conformal blocks, the AGT dual to $\Ztop$, are analytic functions, away from poles and branchcuts, in all of their parameters. In studying how to recover the topological string expansions from conformal field theory, we can thus hope to learn how to move beyond these formal expansions.

To drop the caveat {\it in the field theory limit}, the correspondence needs to be extended to a q-deformed version of conformal field theory, based on a deformed Virasoro algebra \cite{Shiraishi:1995rp}, as initiated in \cite{Awata:2009ur,Awata:2010yy,Taki:2014fva}.

In the papers \cite{KashaniPoor:2012wb,Kashani-Poor:2013oza}, Troost and I studied two instances of the AGT correspondence in detail, corresponding to $\cN=2^*$ and $N_f=4$ Seiberg-Witten theory. The dual conformal field theory quantities are the one-point block on the torus and the four-point block on the sphere. Our principal goals were to recover the genus expansion of the topological string partition function from the conformal field theory perspective, and to uncover how quasi-modularity arises in the conformal field theory approach.

\subsection{Relevant notions in conformal field theory}
\subsubsection{Conformal blocks}
Conformal blocks are universal chiral building blocks of conformal field theory $n$-point functions. They are completely determined by the Virasoro algebra. Thus, on the torus, the one-point function
\be \label{one-point}
\langle \Vt_{{h}_m} \rangle_\tau = \Tr \, \Vc_{h_m} q^{L_0 - \frac{c}{24}} \bar{q}^{\bar{L}_0 - \frac{c}{24}} 
\ee 
can be expressed as a sum over holomorphically factorized contributions 
\be \label{fact_torus}
\langle \Vt_{{h}_m} \rangle_\tau =\sum_{{h}} C^{h}_{{{h}_m} ,{h}} (q \bar{q})^{{h}-\frac{c}{24}}|{\cal F}_{{h}_m} ^{h} (q)|^2 \,.
\ee
The conformal blocks ${\cal F}_{{h}_m} ^{h} (q)$ are meromorphic functions of the Teichm\"uller parameter $\tau$ of the one-punctured torus (by translation invariance, the position of the insertion is irrelevant), as well as of the weight $h_m$ of the insertion, and the summation parameter $h$, the so-called intermediate weight. This sum extends over all primary weights occurring in the trace (\ref{one-point}). The sum over all descendants is incorporated in the respective conformal blocks.\footnote{We will recall the notion of primary and descendant states in the next subsection.} Note that the dynamical information of the conformal field theory is encoded purely in the three-point function $C^{h}_{{{h}_m} ,{h}}$, which in particular determines the range of the sum over intermediate weights. 

To obtain the corresponding expression for the four-point function on the sphere,
\be   \label{four-point}
C_4  = \langle V_{h_1}(z_1) V_{h_2}(z_2) V_{h_3}(z_3) V_{h_4}(z_4) \rangle \,, 
\ee 
we first use conformal invariance to map three of the four insertion points to $0$, $1$, and $\infty$ respectively. 
%The required conformal transformation is
%\be 
%z' = \frac{(z-z_4)(z_2-z_1)}{(z-z_1)(z_2-z_4)}  \,.   \label{coordinatetransformation}
%\ee
The final point is then mapped to the so-called cross-ratio
\be 
x = \frac{(z_3-z_4)(z_2-z_1)}{(z_3-z_1)(z_2-z_4)} \,, 
\ee
which can serve as a coordinate on the Teichm\"uller space of the four-punctured sphere. By performing this transformation, we can express $C_4$ as
\ba
C_4 &=& \left| \frac{(z_4-z_1)(z_2-z_1)}{z_2-z_4} \right|^{2\sum h_i} \prod_{i \neq 1} |z_i-z_1|^{-4h_i} \lim_{z \rightarrow z_1} |z-z_1|^{-4 h_1} \langle V_{h_1}(z'(z))  V_{h_2}(1) V_{h_3}(x) V_{h_4}(0) \rangle \nn \\
&=&\left| \frac{(z_4-z_1)(z_2-z_1)}{z_2-z_4} \right|^{2\sum h_i-4h_1} \prod_{i \neq 1} |z_i-z_1|^{-4h_i} \,G_{1234}(x)  \,. 
\ea
The quantity $G$, defined as
\be
G_{1234}(x) = \lim_{z \rightarrow z_1} |(z'(z)|^{4h_1} \langle V_{h_1}(z'(z))  V_{h_2}(1) V_{h_3}(x) V_{h_4}(0) \rangle = \langle h_{1} | V_{h_2}(1) V_{h_3} (x) | h_4 \rangle \,,
\ee
is then decomposed analogously to (\ref{fact_torus}) as
\be  \label{fact_sphere}
G_{1234}(x) = \sum_h C_{12 h} C_{34}^{h}  \cF_{14}^{23}(h|x) \bar{\cF}_{14}^{23}(h|\bar{x})  \,.
\ee
The conformal blocks $\cF_{14}^{23}(h|x)$ again depend meromorphically on the Teichm\"uller parameter, the weights of all insertions $h_1, \ldots, h_4$, and the intermediate weight $h$. The sum over $h$ here arises upon considering the operator product expansion of $V_{h_3}(x)$ and $V_{h_4}(0)$. It hence extends over all primary states occurring in this OPE.

From the derivation of the holomorphic factorization of $n$-point functions \cite{Belavin:1984vu}, the computation of conformal blocks order by order in power of the Teichm\"uller parameter is a straightforward but tedious algorithmic exercise. This computation can be simplified by relying on recursion relations \cite{Zamo1987bis,Zamolodchikov:1990ww} satisfied by the blocks. In the current context of relating to the genus expansion of the instanton partition function, such recursions were studied in \cite{KashaniPoor:2012wb}. Here, we wish to discuss a method which permits the computation of all order results in the Teichm\"uller parameter, and thus permits reproducing the quasi-modular results of section \ref{s:hae_ir}. This method relies on imposing null vector decoupling on the conformal field theory correlators.

\subsubsection{Null vector decoupling}
The sums that appear in the equations (\ref{fact_torus}) and (\ref{fact_sphere}) are over the weights of primary states in the spectrum of the conformal field theory in question. A primary state $| h \rangle$ is an eigenstate of the Virasoro generator $L_0$, $L_0 |h \rangle = h |h \rangle$, that is annihilated by all positive mode Virasoro generators, $L_n |h \rangle = 0$ for $n > 0$. Descendant states are obtained by acting with an arbitrary number of negative mode Virasoro generators $L_n$, $n < 0$, on the primary state, $\prod L_{-k_i} | h \rangle$.  The sum $l= \sum_i k_i$ is the level of the descendant, its weight is easily seen to be $h+l$. The vector space spanned by a primary state together with all of its descendants is referred to as a Verma module. It furnishes a representation space of the Virasoro algebra. If a descendant of the primary is itself primary, i.e. is annihilated by all positive mode Virasoro generators, then this representation is reducible. Such descendant states are referred to as null vectors. Primary states that exhibit null vectors in their Verma modules are called degenerate. They have been classified. They occur in a family parametrized by two positive integers $m$ and $n$, and their weights are given by
\be 
h_{mn} = \frac{Q^2}{4} - \frac{1}{4} (m b + n \frac{1}{b})^2 \,.
\ee 
The parameter $Q = b + \frac{1}{b}$ is related to the central charge $c$ of the Virasoro algebra via $c = 1 + 6 Q^2$. Correlators involving null vectors can consistently be set to zero. This is referred to as null vector decoupling. Setting these correlators to zero will not modify correlators with states of smaller level, as any attempt to lower the level of the null vector by acting on it with a positive mode Virasoro generator will annihilate it.

As the Virasoro generators act within correlators as differential operators, correlators with degenerate insertions satisfy differential equations, called null vector decoupling equations. The order of the differential equation coincides with the level of the null vector. The null vector decoupling equation of the four-point function with an additional degenerate insertion $V_{(2,1)}$ at level 2 is immediate to write down,
\be  \label{four_pt_NVD}
\left[ \partial_z^2 +b^2 \left( \sum_{k=0}^3 \frac{h_k}{(z-z_k)^2} + \frac{ \partial_k}{z-z_k} \right) \right] \langle V_{(2,1)}(z) V_{\ha} (z_0) \ldots V_{\hd}(z_3) \rangle = 0 \,.
\ee
The analogous equation for the one-point function requires invoking the conformal Ward identity on the torus \cite{Eguchi:1986sb}. It is given by
\be  \label{one_pt_NVD}
\Big[ \frac{1}{b^2} \partial_z^2 +(  2 \eta_1 z -\zeta(z)   )\partial_{z} + 2 \pi i \partial_\tau + 2 h_{21} \eta_1 +  {h}_{m} ( \wp (z) + 2 \eta_1) \Big] Z \langle \Vt_{(2,1)} (z) \Vt_{{h}_m} (0) \rangle =  0 \,.  
\ee
Here, $Z$ is the partition function on the torus, $\wp$ is the Weierstrass $\wp$-function, $\zeta$ is its primitive, $\zeta'(z) = - \wp(z)$, and $\eta_1 = \frac{\pi^2}{6} E_2$.

These differential equations are satisfied by the full correlator. To isolate the contribution from a conformal block of a given intermediate weight, we need to impose the appropriate boundary conditions on the solution: the monodromy of the solution as the insertion point of the degenerate operator circles the origin, or the origin and $x$, in the case of the torus and the spherical block respectively. This is determined by considering the operator product expansion of the degenerate operator with the operator $V_h$, which, as described above, occurs either directly in the trace (\ref{one-point}), or in the OPE of $V_{h_3}$ and $V_{h_4}$ when evaluating (\ref{four-point}):
\ba  \label{OPE}
V_{(2,1)}(z) V_{h}(x) & =& C_{h_{21} , h}^{h_{+}} (z-x)^{h_{+} - h_{21} - h} \left( V_{h_+}(x) + \ldots \right)  \\
&&+ \,C_{h_{21} , h}^{h_{-}} (z-x)^{h_{-} - h_{21} - h} \left( V_{-}(x) + \ldots \right) \,.   \nn
\ea 
Note that the OPE of the degenerate operator $V_{(2,1)}$ with any other operator involves only two primaries. Using the standard parametrization of weights in terms of Liouville momenta $\alpha$,
\be  \label{def_L_mom}
h = \alpha (Q - \alpha) \,,
\ee
the Liouville momenta of these primaries are given by $\alpha_\pm = \alpha \pm \frac{b}{2} $. The $\ldots$ in the parentheses above indicate a power series in $z$. The monodromy in $z$ is hence determined entirely by the exponent $h_{\pm} - h_{21} - h$.

The presentation up to this point is exact. What we have arrived at is a differential equation and boundary conditions to determine the two-point block on the torus and the five-point block on the sphere respectively. To extract from these results the conformal blocks of interest, the contribution of the degenerate operator needs to be identified. This is only possible in a limit of parameter space where the ratio of the weights of the original insertions to that of the degenerate operator goes to infinity, as we discuss below in subsection \ref{s:fact_limit}.

\newpage

\subsection{The genus expansion and modularity}
\subsubsection{The AGT dictionary}
Having reviewed the necessary notions of conformal field theory, we are now ready to present the AGT dictionary: the weights of the insertions of the conformal blocks correspond to the masses of the flavors of the gauge theory, and the intermediate weights to the vacuum expectation values of the vector multiplet scalars. The Teichm\"uller parameters on which the conformal blocks depend map to coordinates on the space of marginal couplings of the gauge theory (i.e. ultraviolet couplings). Finally, the central charge of the conformal field theory can be expressed in terms of the $\epsilon$-parameters of the gauge theory, which we package into the two couplings $g_s^2= \epsilon_1 \epsilon_2$ and $s=(\epsilon_1+\epsilon_2)^2$. In formulae,
\be \label{dictionary}
b^2 = \frac{\epsilon_2}{\epsilon_1}  \,,  \quad
h_m = \frac{Q^2}{4} - \frac{m^2}{\epsilon_1 \epsilon_2} \,,  \quad 
h = \frac{Q^2}{4} - \frac{a^2}{\epsilon_1 \epsilon_2} \,.
\ee
As the $\epsilon$-parameters also enter in the dictionary relating weights on the one side of the correspondence to masses and vacuum expectation values on the other, the small $\epsilon$ limit, which we will refer to as the genus expansion limit, implies taking all weights to be large. Note that this does not yet imply a particular limit for $b$, hence for the central charge $c$.

\subsubsection{The factorization limit}  \label{s:fact_limit}
For the null vector decoupling equations to be useful for our purposes, we need to be able to extract the blocks of interest from the solutions of these equations, which yield these blocks with the additional insertion of a degenerate field. This can be achieved, but at the price of taking the semi-classical limit $b \rightarrow 0$, equivalent via (\ref{dictionary}) to $\epsilon_2 \ll \epsilon_1$. In this limit, one can argue using the tools of Liouville theory (though the result must be general, as conformal blocks are universal objects) that conformal blocks factorize when one considers heavy ($\alpha \xrightarrow[b \rightarrow 0]{} \infty$) and light ($\alpha \xrightarrow[b \rightarrow 0]{} 0$) insertions simultaneously (see e.g. \cite{Harlow:2011ny} for a lucid exposition). The intuition behind this factorization is that in the semi-classical limit, heavy insertions backreact on the classical Liouville metric, while the contribution of light insertions can be approximated by multiplicative factors.

The parametrization  (\ref{dictionary}) of the weights of the insertions implies that at finite $m$, these scale towards infinity in the semi-classical limit. By contrast, the Liouville momentum $\alpha_{(2,1)} = -b/2$ of the degenerate operator $V_{(2,1)}$ identifies it as a light insertion. We will thus be able to factor out its contribution to the block semi-classically.

\newpage

\subsubsection{The torus one-point block} \label{s:torus_one_pt}
Let us now apply the computational strategy outlined above to the torus one-point block. With the ansatz
\be
 Z  \langle \Vt_{(2,1)}(z) \Vt_{{h}_m} (0) \rangle_\tau = \theta_1(z|\tau) ^{\frac{b^2}{2}} \eta(\tau)^{2({h}_m - b^2 -1)} \Psi(z|\tau) \,,
\label{diffeqansatz}
\ee
the null vector decoupling equation (\ref{one_pt_NVD}) takes the form of a holomorphic Schr\"odinger equation:
\be \label{The_de}
\left[- \partial_z^2 - \big(\frac{1}{4} - \frac{1}{\epsilon_1^2} m^2\big) \wp(z) \right] \Psi(z|\tau) = \frac{\epsilon_2}{\epsilon_1}
2 \pi i \partial_\tau \Psi(z|\tau) \,.  
\ee 
In accord with the factorization property in the semi-classical limit,
\be
\langle V_{(2,1)}(z,\bar{z}) V_{h_m}(0) \rangle  \approx  
e^{-\frac{1}{2} \phi_{cl}(z, \bar{z})}\langle V_{h_m}(0) \rangle \,,
\ee
we make the following ansatz for $\Psi$:
\begin{eqnarray} \label{ansatz_psi}
\Psi(z|\tau) &=& \exp \left[ \frac{1}{\epsilon_1 \epsilon_2}  {\cal F}(\tau) + \frac{1}{\epsilon_1} {\cal W}(z|\tau)  \right] \,.
\end{eqnarray}
The boundary condition which follows from the discussion around (\ref{OPE}) now translates into
 \be 
\cW(z+1) - \cW(z) = \pm 2\pi i a \,,
\ee
and the solution of the differential equation with the boundary condition imposed will yield the sought after conformal block $\exp \frac{1}{\epsilon_1 \epsilon_2}  {\cal F}(\tau) $. 

To solve the equation, we make the formal ansatz
\be
 {\cal F}(\tau) =
\sum_{n=0}^\infty {\cal F}_n(\tau) \epsilon_1^n \,, \quad {\cal  W}(z|\tau) = \sum_{n=0}^\infty {\cal W}_n(z|\tau) \epsilon_1^n \,.   \label{pert_ansatz}
\ee 
The AGT correspondence predicts $F^{(n,0)} = \cF_{2n}$.
Note that introducing a second expansion in $\epsilon_2$ would not longer permit the distinction between $\cF$ and $\cW$ based on the leading $\epsilon_2$ behavior exhibited in (\ref{ansatz_psi}). Until we find an alternative criterium for separating the two, we must hence work in the $g_s=0$ limit. 

We arrive at the following system of equations to determine $\cF_n$ and $\cW_n$:
\ba \label{NVD_system}
-{\cW'_0}^2 + m^2 \wp &=& (2 \pi i)^2 q \partial_q \cF_0 \,,  \\
-\cW''_0 - 2 \cW'_0 \cW'_1&=& (2 \pi i)^2 q \partial_q \cF_1 \,,  \nn \\
- \cW''_1 - {\cW'_1}^2  -2 \cW'_0 \cW'_2 - \frac{1}{4} \wp(z) &=&  (2 \pi i)^2 q \partial_q \cF_2 \,,\nn\\
- \cW''_{n} - \sum_{i=0}^{n+1} \cW'_i \cW'_{n+1-i} &=& (2 \pi i)^2 q \partial_q \cF_{n+1}  \quad \mbox{for} \quad {n \ge 2} \,, \nn
\ea
together with the boundary condition
\be
\oint_A \cW_0' = \pm 2 \pi i a \,, \quad \oint_A \cW_i' = 0  \quad \mbox{for} \quad i>0
\,,  \label{bc} 
\ee
where the subscript $A$ indicates the integral over the $A$-cycle of the torus defined by the lattice spanned by the lattice vectors $(0,1)$ and $(0,\tau)$.

The recursive solution of these equations in a $\frac{m}{a}$ expansion is straightforward. It requires evaluating integrals of the form $\oint_A \wp^n$. These can be evaluated recursively \cite{Halphen,Grosset}. They take values in the ring $\widetilde{M}(SL(2,\IZ))$ of quasi-modular forms encountered in subsection \ref{s:modularity}. The computation thus naturally leads to modular expressions for $\partial_\tau \cF_n$. $\widetilde{M}(SL(2,\IZ))$ is not closed under integration however. For its elements to integrate to quasi-modular forms, the coefficients must satisfy algebraic constraints which follow from the Ramanujan identities. In \cite{KashaniPoor:2012wb}, we verified experimentally that these constraints are met up to a given order of computation. The proof of this property \cite{Kashani-Poor:2014mua} follows upon uncovering the special geometry underlying the equation (\ref{The_de}), as we outline in section \ref{quantum} below. To give a flavor of the results one encounters, let us reproduce the following few amplitudes here:
\be
\cF_2 = - \frac{\log \eta}{2} - \frac{E_2}{2^53} \frac{m^2}{a^2} + O((\frac{m^2}{a^2})^2)  \,, 
\ee
\be
\cF_4 = \frac{E_2}{2^8 3 a^2} + \frac{(5 E_2^2 + 9 E_4)}{2^9 3 \cdot5 a^2}\frac{m^2}{a^2} + O((\frac{m^2}{a^2})^2)  \,, \nn
\ee
\be
\cF_6 = - \frac{5 E_2^2 + 13 E_4}{2^{13} 3^2 5 a^4} - \frac{(35 E_2^3 + 168 E_2 E_4 + 355 E_6)}{2^{14} 3^4 7 a^4} \frac{m^2}{a^2} + O((\frac{m^2}{a^2})^2) \,. \nn
\ee
Up to a rescaling of $a$, these coincide with the results (\ref{results_N4_hae}) obtained via the holomorphic anomaly equations. 

\subsubsection{The spherical four-point block}  \label{s:four_pt}
The occurrence of modularity in the computation of the one-point toric block is perhaps ultimately not surprising (though it had not been observed prior to \cite{KashaniPoor:2012wb}), given the torus underlying the problem. Indeed, as we witnessed above, quasi-modular forms arise upon integration of the Weierstrass $\wp$-function, in terms of which the Ward identities on a torus are naturally formulated. The occurrence of modularity in the spherical four-point block might naively appear less obvious. However, when formulated in the correct variables, the computation of the four-point block is in fact very similar to the toric case.

We can motivate the appropriate choice of variables as follows. Above, we introduced the cross ratio $x$ as a representative of the class of four insertion points $(z_1,z_2,z_3,z_4)$ acted upon by global conformal transformations, and expressed the spherical block as a function of $x$. This variable takes values on the three punctured sphere. To move the punctures to a more convenient location, we can instead consider the parametrization
\be  \label{cross}
x = \frac{e_3-e_2}{e_3-e_1}(\tau) \,,
\ee
with the $e_i$ the half-periods of the Weierstrass $\wp$-function already encountered above. The right-hand side is invariant under the action of $\Gamma(2) \subset SL(2,\IZ)$ introduced in subsection \ref{s:s-duality}. We have hence mapped the moduli space to $\tau \in \IH / \Gamma(2)$, and the punctures to the cusps of this domain. The permutations of the insertion points $z_i$, which act as rational transformations on $x$, are realized by the action of $S_3 \cong SL(2,\IZ) / \Gamma(2)$ on $\tau$. From these considerations, it is natural to expect modular behavior  of the spherical conformal block under $\Gamma(2)$ when it is expressed as a function of $\tau$. 

Motivated by (\ref{cross}), we introduce a 2:1 map from the torus to the sphere via \cite{P}
\be
z = \frac{\wp(u) - e_3}{e_1 - e_3} \,.
\ee
In terms of this coordinate, the null vector decoupling equation (\ref{four_pt_NVD}) takes the form \cite{Fateev:2009me}
\be  \label{null_vect_ell}
\left(\partial_u^2 + 4 b^2 \sum_{i=0}^3 \hat{h}_i \wp(u+ \omega_i) \right) \Psi(u|\tau) =  - 4 \pi i b^2 \partial_\tau  \Psi (u|\tau)    \,,
\ee
with $\omega_i$ the two-torsion points of the torus introduced in subsection \ref{s:n2_star_curve}, and 
\be
\hat{h}_i = h_i - \frac{b^2}{4} - \frac{3}{16 b^2} - \frac{1}{2} \,.
\ee
$\Psi(u|\tau)$ is related to the five-point correlator via a somewhat complicated prefactor \cite{Kashani-Poor:2013oza}, which however modifies the semi-classical expansion of the conformal block only at leading and subleading order in $\epsilon_1$. Comparing to (\ref{The_de}), the similarity of this problem to the computation of the one-point block is manifest. Computationally, a slight generalization of the period integrals of powers of the Weierstrass $\wp$-function is required \cite{Kashani-Poor:2013oza}. This again leads to expressions for $\cF_n$ in terms of quasi-modular forms that coincide with the computations presented in the previous section based on the holomorphic anomaly.

\subsubsection{Seiberg-Witten and quantum geometry from null vector decoupling} \label{quantum}
The results we obtained above are modular in the Teichm\"uller parameter of the punctured Riemann surface associated to the conformal block in question. By the AGT dictionary, this parameter corresponds to the ultraviolet coupling of the dual gauge theory. This modularity is thus a reflection of the S-duality of the gauge theory as described in subsection \ref{s:s-duality} above. The modularity underlying the exact results of section \ref{s:n2_holanomaly} in contrast is based on the monodromy group of the gauge theory, and the corresponding modular parameter is the infrared coupling of the gauge theory. In this section, we want to identify this parameter in conformal field theory \cite{Kashani-Poor:2014mua}. We will argue within the context of $\cN=2^*$ theory.

The entry point once again is the system of equations (\ref{NVD_system}) derived from null vector decoupling. Reassembling the expansion coefficients, the boundary condition (\ref{bc}) can be expressed as 
\be   \label{genper}
\int_0^1 \sqrt{m^2 \wp - (2 \pi i)^2 q\partial_q \cF - \epsilon_1 \cW'' -  \epsilon_1^2\, \frac{\wp}{4} } \,dz= \pm 2 \pi i a  \,.
\ee
This and all following equalities involving $\cF$ and $\cW$ are to be interpreted in the sense of equalities of formal power series. We show that the integrand 
\be
\lambda = \sqrt{m^2 \wp - (2 \pi i)^2 q\partial_q \cF - \epsilon_1 \cW'' -\epsilon_1^2\, \frac{\wp}{4} } \, dz
\ee
can be interpreted as a deformed Seiberg-Witten differential, in that it satisfies the $\epsilon$-deformed special geometry relation
\be  \label{deformed_special_geometry}
\oint_B \lambda = -\frac{1}{2}\frac{\partial \cF}{\partial a}  \,.
\ee
The proof of this relation follows along the lines of the proof of the Riemann bilinear identity, upon the correct choice of formal differential form. $\epsilon$-deformed special geometry is sometimes referred to as quantum geometry.

The Riemann surface on which the integrals are to be performed follows by considering the leading $\epsilon_1$ contribution
\be \label{lambda0}
\lambda_0 = \sqrt{ m^2 \wp - u} \,dz \,, \quad u := 2\pi i \, \partial_\tau \cF_0  
\ee
to $\lambda$. The square root is single-valued on the genus 2 surface defined by
\be  \label{genus_2_sw}
t^2 = m^2 \wp - u  \,,
\ee
the double cover of the definition domain of $\wp$. Two holomorphic one-forms exist on this surface. One yields the ultraviolet coupling of the gauge theory as the ratio of its $B$ to its $A$ period (relative to one sheet, the periods on the second sheet merely differ in sign), the other the infrared coupling. The relation of the curve (\ref{genus_2_sw}) to $\cN=2^*$ theory was already pointed out in \cite{DW}. We thus reproduce all of the Seiberg-Witten data from within conformal field theory.

Finally, notice that the relation (\ref{deformed_special_geometry}) provides the missing argument proving the quasi-modularity of $\cF$, as the period integrals of $\lambda$ are manifestly quasi-modular.

\newpage

\section{Perspectives}
As promised in the introduction, we have seen that the passage from the physical to the topological string enhances our command of the theory dramatically, giving rise to a variety of computational techniques leading to all order results in $g_s$ or the K\"ahler parameters for $\Ztop$. The AGT correspondence presents a promising avenue towards studying the nature of the remaining series expansions, and the analytic properties of $\Ztop$. Much work remains to be done: to elevate the results of the previous section past $g=0$, to move away from the field theory limit, to generalize away from geometric engineering geometries. At the end of this path, a $q$-deformed version of conformal field theory beckons that has $\Ztop$ on arbitrary toric geometries as its observables. Whether this will shed light on the ultimate goal of finding a non-perturbative completion of perturbative string theory remains to be seen, but we are guaranteed to encounter much intricate and beautiful structure along the way.

\hspace{5cm}

\section*{Acknowledgments}
I would like to thank all of my colleagues with whom I have had the good fortune to collaborate and discuss physics over the last fifteen years, the members of my thesis committee for agreeing to the task, and, as always, my family for their love and support.

\newpage

\bibliography{biblio_hab}
\bibliographystyle{utcaps}

\end{document}